\pdfoutput=1
\documentclass[12pt,a4paper]{article}
\usepackage{ifthen} 
\newboolean{pdflatex}
\setboolean{pdflatex}{true} 

\newboolean{articletitles}
\setboolean{articletitles}{true} 

\newboolean{uprightparticles}
\setboolean{uprightparticles}{false} 

\newboolean{inbibliography}
\setboolean{inbibliography}{false} 

\usepackage[top=1in, bottom=1.25in, left=1in, right=1in]{geometry}

\usepackage{microtype}
\usepackage{lineno}  
\usepackage{xspace} 
\usepackage{caption} 

\usepackage{graphicx}  
\usepackage{color}
\usepackage{colortbl}

\usepackage{amsmath} 
\usepackage{amssymb}
\usepackage{amsfonts}
\usepackage{upgreek} 

\newcommand*\patchAmsMathEnvironmentForLineno[1]{%
\expandafter\let\csname old#1\expandafter\endcsname\csname #1\endcsname
\expandafter\let\csname oldend#1\expandafter\endcsname\csname
end#1\endcsname
 \renewenvironment{#1}%
   {\linenomath\csname old#1\endcsname}%
   {\csname oldend#1\endcsname\endlinenomath}%
}
\newcommand*\patchBothAmsMathEnvironmentsForLineno[1]{%
  \patchAmsMathEnvironmentForLineno{#1}%
  \patchAmsMathEnvironmentForLineno{#1*}%
}
\AtBeginDocument{%
\patchBothAmsMathEnvironmentsForLineno{equation}%
\patchBothAmsMathEnvironmentsForLineno{align}%
\patchBothAmsMathEnvironmentsForLineno{flalign}%
\patchBothAmsMathEnvironmentsForLineno{alignat}%
\patchBothAmsMathEnvironmentsForLineno{gather}%
\patchBothAmsMathEnvironmentsForLineno{multline}%
\patchBothAmsMathEnvironmentsForLineno{eqnarray}%
}

\usepackage{hyperref}    
\usepackage[all]{hypcap} 

\usepackage{xspace} 
\usepackage{upgreek}

\def\lhcb {\mbox{LHCb}\xspace}

\def\MagUp {\mbox{\em Mag\kern -0.05em Up}\xspace}

\ifthenelse{\boolean{uprightparticles}}%
{

 \def\Ppi         {\ensuremath{\uppi}\xspace}

 \def\Ppsi        {\ensuremath{\uppsi}\xspace}

 \def\PDelta      {\ensuremath{\Delta}\xspace}                 
 \def\PXi      {\ensuremath{\Xi}\xspace}                 
 \def\PLambda      {\ensuremath{\Lambda}\xspace}                 
 \def\PSigma      {\ensuremath{\Sigma}\xspace}                 
 \def\POmega      {\ensuremath{\Omega}\xspace}                 
 \def\PUpsilon      {\ensuremath{\Upsilon}\xspace}                 
 
 \def\PB      {\ensuremath{\mathrm{B}}\xspace}                 
                  
 \def\PD      {\ensuremath{\mathrm{D}}\xspace}

 \def\PJ      {\ensuremath{\mathrm{J}}\xspace}                 
 \def\PK      {\ensuremath{\mathrm{K}}\xspace}

 \def\Pb      {\ensuremath{\mathrm{b}}\xspace}                 
 \def\Pc      {\ensuremath{\mathrm{c}}\xspace}                 
 \def\Pd      {\ensuremath{\mathrm{d}}\xspace}

 \def\Pi      {\ensuremath{\mathrm{i}}\xspace}

 \def\Pp      {\ensuremath{\mathrm{p}}\xspace}

 \def\Ps      {\ensuremath{\mathrm{s}}\xspace}

}
{

 \def\Ppi         {\ensuremath{\pi}\xspace}

 \def\Ppsi        {\ensuremath{\psi}\xspace}                 
                  
 \mathchardef\PDelta="7101
 \mathchardef\PXi="7104
 \mathchardef\PLambda="7103
 \mathchardef\PSigma="7106
 \mathchardef\POmega="710A
 \mathchardef\PUpsilon="7107
                  
 \def\PB      {\ensuremath{B}\xspace}                 
                  
 \def\PD      {\ensuremath{D}\xspace}

 \def\PJ      {\ensuremath{J}\xspace}                 
 \def\PK      {\ensuremath{K}\xspace}

 \def\Pb      {\ensuremath{b}\xspace}                 
 \def\Pc      {\ensuremath{c}\xspace}                 
 \def\Pd      {\ensuremath{d}\xspace}

 \def\Pi      {\ensuremath{i}\xspace}

 \def\Pp      {\ensuremath{p}\xspace}

 \def\Ps      {\ensuremath{s}\xspace}

}

\makeatletter
\ifcase \@ptsize \relax
  \newcommand{\miniscule}{\@setfontsize\miniscule{4}{5}}
\or
  \newcommand{\miniscule}{\@setfontsize\miniscule{5}{6}}
\or
  \newcommand{\miniscule}{\@setfontsize\miniscule{5}{6}}
\fi
\makeatother

\DeclareRobustCommand{\optbar}[1]{\shortstack{{\miniscule (\rule[.5ex]{1.25em}{.18mm})}
  \\ [-.7ex] $#1$}}

\def\dquark    {{\ensuremath{\Pd}}\xspace}

\def\squark    {{\ensuremath{\Ps}}\xspace}

\def\cquark    {{\ensuremath{\Pc}}\xspace}

\def\bquark    {{\ensuremath{\Pb}}\xspace}

\def\pion   {{\ensuremath{\Ppi}}\xspace}
\def\piz    {{\ensuremath{\pion^0}}\xspace}

\def\pip    {{\ensuremath{\pion^+}}\xspace}
\def\pim    {{\ensuremath{\pion^-}}\xspace}
\def\pipm   {{\ensuremath{\pion^\pm}}\xspace}

\def\kaon    {{\ensuremath{\PK}}\xspace}
  \def\Kbar    {{\kern 0.2em\overline{\kern -0.2em \PK}{}}\xspace}

\def\KorKbar    {\kern 0.18em\optbar{\kern -0.18em K}{}\xspace}
\def\Kz      {{\ensuremath{\kaon^0}}\xspace}

\def\Kp      {{\ensuremath{\kaon^+}}\xspace}
\def\Km      {{\ensuremath{\kaon^-}}\xspace}
\def\Kpm     {{\ensuremath{\kaon^\pm}}\xspace}

\def\KS      {{\ensuremath{\kaon^0_{\mathrm{ \scriptscriptstyle S}}}}\xspace}

\def\Kstarm  {{\ensuremath{\kaon^{*-}}}\xspace}

  \def\Dbar    {{\kern 0.2em\overline{\kern -0.2em \PD}{}}\xspace}
\def\D       {{\ensuremath{\PD}}\xspace}

\def\DorDbar    {\kern 0.18em\optbar{\kern -0.18em D}{}\xspace}
\def\Dz      {{\ensuremath{\D^0}}\xspace}
\def\Dzb     {{\ensuremath{\Dbar{}^0}}\xspace}

\def\B       {{\ensuremath{\PB}}\xspace}
\def\Bbar    {{\ensuremath{\kern 0.18em\overline{\kern -0.18em \PB}{}}}\xspace}

\def\BorBbar    {\kern 0.18em\optbar{\kern -0.18em B}{}\xspace}
\def\Bz      {{\ensuremath{\B^0}}\xspace}
\def\Bzb     {{\ensuremath{\Bbar{}^0}}\xspace}

\def\Bub     {{\ensuremath{\B^-}}\xspace}

\def\Bm      {{\ensuremath{\Bub}}\xspace}
\def\Bpm     {{\ensuremath{\B^\pm}}\xspace}

\def\Bs      {{\ensuremath{\B^0_\squark}}\xspace}
\def\Bsb     {{\ensuremath{\Bbar{}^0_\squark}}\xspace}

\def\jpsi     {{\ensuremath{{\PJ\mskip -3mu/\mskip -2mu\Ppsi\mskip 2mu}}}\xspace}

  \def\Y#1S{\ensuremath{\PUpsilon{(#1S)}}\xspace}

\def\proton      {{\ensuremath{\Pp}}\xspace}

\def\Xires       {{\ensuremath{\PXi}}\xspace}

\def\Lz          {{\ensuremath{\PLambda}}\xspace}
\def\Lbar        {{\ensuremath{\kern 0.1em\overline{\kern -0.1em\PLambda}}}\xspace}
\def\LorLbar    {\kern 0.18em\optbar{\kern -0.18em \PLambda}{}\xspace}

\def\Omegares    {{\ensuremath{\POmega}}\xspace}

\def\Lb      {{\ensuremath{\Lz^0_\bquark}}\xspace}

\def\Xib     {{\ensuremath{\Xires_\bquark}}\xspace}
\def\Xibz    {{\ensuremath{\Xires^0_\bquark}}\xspace}
\def\Xibm    {{\ensuremath{\Xires^-_\bquark}}\xspace}

\def\Xicz    {{\ensuremath{\Xires^0_\cquark}}\xspace}

\def\Omegab    {{\ensuremath{\Omegares^-_\bquark}}\xspace}

\def\BF         {{\ensuremath{\mathcal{B}}}\xspace}

\def\to                 {\ensuremath{\rightarrow}\xspace}

\def\CP                {{\ensuremath{C\!P}}\xspace}

\def\AT#1     {\ensuremath{A_{\mathrm{T}}^{#1}}\xspace}           

\def\C#1      {\ensuremath{\mathcal{C}_{#1}}\xspace}                       
\def\Cp#1     {\ensuremath{\mathcal{C}_{#1}^{'}}\xspace}                    
\def\Ceff#1   {\ensuremath{\mathcal{C}_{#1}^{\mathrm{(eff)}}}\xspace}        
\def\Cpeff#1  {\ensuremath{\mathcal{C}_{#1}^{'\mathrm{(eff)}}}\xspace}       
\def\Ope#1    {\ensuremath{\mathcal{O}_{#1}}\xspace}                       
\def\Opep#1   {\ensuremath{\mathcal{O}_{#1}^{'}}\xspace}                    

\newcommand{\tev}{\ifthenelse{\boolean{inbibliography}}{\ensuremath{~T\kern -0.05em eV}\xspace}{\ensuremath{\mathrm{\,Te\kern -0.1em V}}}\xspace}
\newcommand{\gev}{\ensuremath{\mathrm{\,Ge\kern -0.1em V}}\xspace}
\newcommand{\mev}{\ensuremath{\mathrm{\,Me\kern -0.1em V}}\xspace}
\newcommand{\kev}{\ensuremath{\mathrm{\,ke\kern -0.1em V}}\xspace}
\newcommand{\ev}{\ensuremath{\mathrm{\,e\kern -0.1em V}}\xspace}
\newcommand{\gevc}{\ensuremath{{\mathrm{\,Ge\kern -0.1em V\!/}c}}\xspace}
\newcommand{\mevc}{\ensuremath{{\mathrm{\,Me\kern -0.1em V\!/}c}}\xspace}
\newcommand{\gevcc}{\ensuremath{{\mathrm{\,Ge\kern -0.1em V\!/}c^2}}\xspace}
\newcommand{\gevgevcccc}{\ensuremath{{\mathrm{\,Ge\kern -0.1em V^2\!/}c^4}}\xspace}
\newcommand{\mevcc}{\ensuremath{{\mathrm{\,Me\kern -0.1em V\!/}c^2}}\xspace}

\def\invfb   {\ensuremath{\mbox{\,fb}^{-1}}\xspace}

\newcommand{\stat}{\ensuremath{\mathrm{\,(stat)}}\xspace}
\newcommand{\syst}{\ensuremath{\mathrm{\,(syst)}}\xspace}

\newcommand{\chisq}{\ensuremath{\chi^2}\xspace}

\newcommand{\chisqip}{\ensuremath{\chi^2_{\text{IP}}}\xspace}

\def\gsim{{~\raise.15em\hbox{$>$}\kern-.85em
          \lower.35em\hbox{$\sim$}~}\xspace}
\def\lsim{{~\raise.15em\hbox{$<$}\kern-.85em
          \lower.35em\hbox{$\sim$}~}\xspace}

\def\sPlot{\mbox{\em sPlot}\xspace}

\def\pt         {\mbox{$p_{\mathrm{ T}}$}\xspace}

\def\tell1  {TELL1\xspace}
\def\ukl1   {UKL1\xspace}

\newcommand{\ie}{\mbox{\itshape i.e.}\xspace}

\usepackage{cite} 
\usepackage{mciteplus}
\allowdisplaybreaks 

\begin{document}

\renewcommand{\thefootnote}{\fnsymbol{footnote}}
\setcounter{footnote}{1}

\begin{titlepage}
\pagenumbering{roman}

\vspace*{-1.5cm}
\centerline{\large EUROPEAN ORGANIZATION FOR NUCLEAR RESEARCH (CERN)}
\vspace*{1.5cm}
\noindent
\begin{tabular*}{\linewidth}{lc@{\extracolsep{\fill}}r@{\extracolsep{0pt}}}
\ifthenelse{\boolean{pdflatex}}
{\vspace*{-2.7cm}\mbox{\!\!\!\includegraphics[width=.14\textwidth]{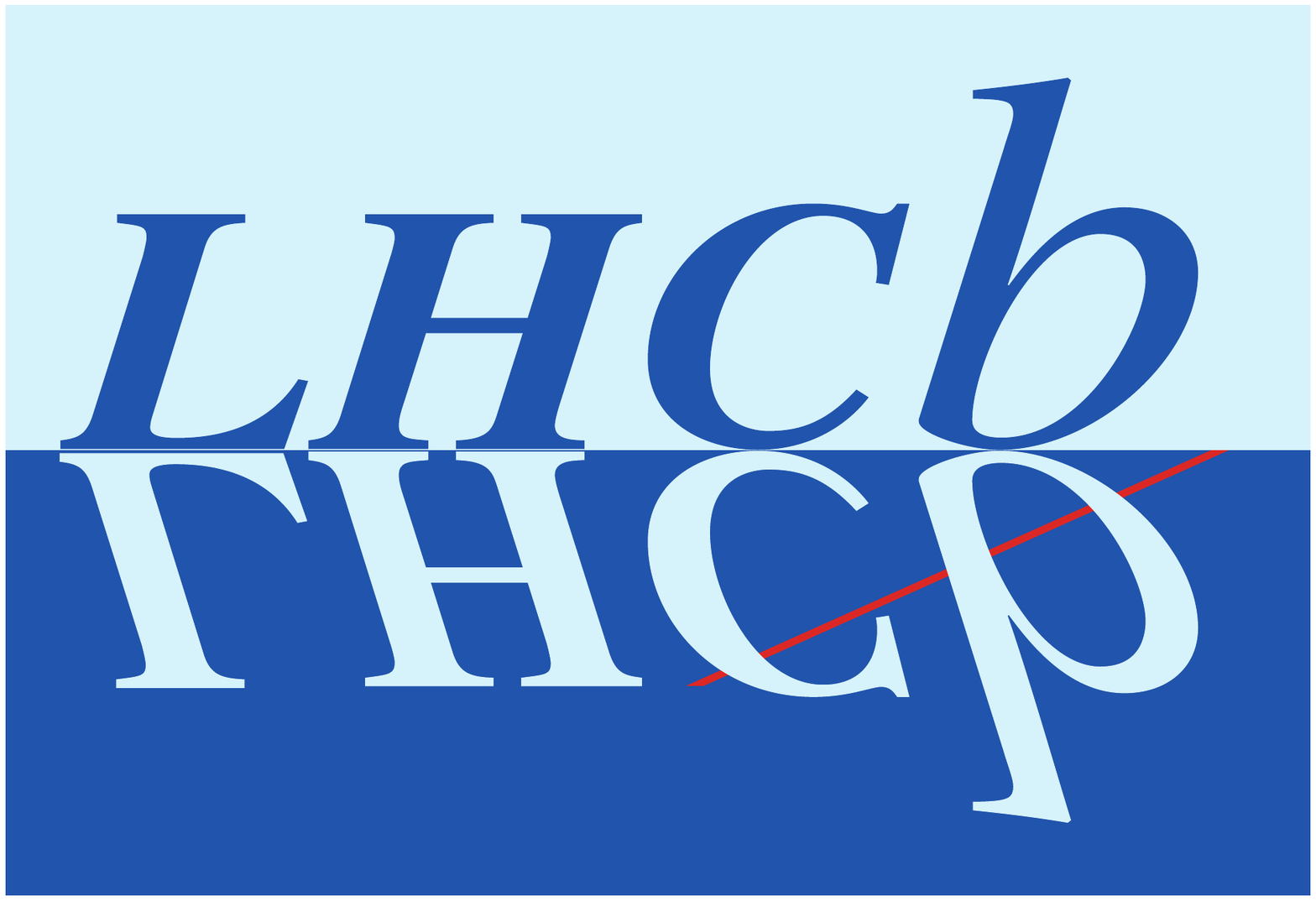}} & &}%
{\vspace*{-1.2cm}\mbox{\!\!\!\includegraphics[width=.12\textwidth]{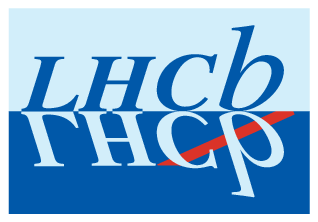}} & &}%
\\
 & & CERN-EP-2016-294 \\  
 & & LHCb-PAPER-2016-050 \\  
 & & December 7, 2016\\ 
 & & \\
\end{tabular*}

\vspace*{4.0cm}

{\normalfont\bfseries\boldmath\huge
\begin{center}
  Observation of the decay $\Xibm \to \proton \Km\Km$
\end{center}
}

\vspace*{2.0cm}

\begin{center}
The LHCb collaboration\footnote{Authors are listed at the end of this Letter.}
\end{center}

\vspace{\fill}

\begin{abstract}
  \noindent
  Decays of the $\Xibm$ and $\Omegab$ baryons to the charmless final states $\proton h^{-}h^{\prime -}$, where $h^{(\prime)}$ denotes a kaon or pion, are searched for with the LHCb detector.
  The analysis is based on a sample of proton-proton collision data collected at centre-of-mass energies $\sqrt{s} = 7$ and $8 \tev$, corresponding to an integrated luminosity of $3 \invfb$.
  The decay $\Xibm \to \proton \Km\Km$ is observed with a significance of $8.7$ standard deviations, and evidence at the level of $3.4$ standard deviations is found for the $\Xibm \to \proton \Km\pim$ decay.
  Results are reported, relative to the $\Bm\to \Kp\Km\Km$ normalisation channel, for the products of branching fractions and \bquark-hadron production fractions.
  The branching fractions of $\Xibm \to \proton \Km\pim$ and $\Xibm \to \proton \pim\pim$ relative to $\Xibm \to \proton \Km\Km$ decays are also measured.
\end{abstract}

\vspace*{2.0cm}

\begin{center}
  Published in Phys.~Rev.~Lett.
\end{center}

\vspace{\fill}

{\footnotesize 
\centerline{\copyright~CERN on behalf of the \lhcb collaboration, licence \href{http://creativecommons.org/licenses/by/4.0/}{CC-BY-4.0}.}}
\vspace*{2mm}

\end{titlepage}

\newpage
\setcounter{page}{2}
\mbox{~}

\cleardoublepage

\renewcommand{\thefootnote}{\arabic{footnote}}
\setcounter{footnote}{0}

\pagestyle{plain} 
\setcounter{page}{1}
\pagenumbering{arabic}

Decays of \bquark hadrons to final states that do not contain charm quarks provide fertile ground for studies of \CP violation, \ie\ the breaking of symmetry under the combined charge conjugation and parity operations.
Significant asymmetries have been observed between \B and \Bbar partial widths in \mbox{$\Bzb \to \Km\pip$}~\cite{Lees:2012mma,Duh:2012ie,Aaltonen:2014vra,LHCb-PAPER-2013-018} and \mbox{$\Bsb \to \Kp\pim$}~\cite{Aaltonen:2014vra,LHCb-PAPER-2013-018} decays.
Even larger \CP-violation effects have been observed in regions of the phase space of \mbox{$\Bm \to \pip\pim\pim$}, $\Km\pip\pim$, $\Kp\Km\Km$ and $\Kp\Km\pim$ decays~\cite{LHCb-PAPER-2013-027,LHCb-PAPER-2013-051,LHCb-PAPER-2014-044}.
A number of theoretical approaches~\cite{Gronau:2008gu,Baek:2009hv,Ciuchini:2006kv,Ciuchini:2006st,Gronau:2006qn,Gronau:2007vr,Bediaga:2006jk,Gronau:2010dd,Gronau:2010kq,Imbeault:2010xg,Bhattacharya:2015uua} have been proposed to determine whether the observed effects are consistent with being solely due to the non-zero phase in the quark mixing matrix~\cite{Cabibbo:1963yz,Kobayashi:1973fv} of the Standard Model, or whether additional sources of asymmetry are contributing.

Breaking of the symmetry between matter and antimatter has not yet been observed with a significance of more than five standard deviations ($\sigma$) in the properties of any baryon.  
Recently, however, the first evidence of \CP violation in the \bquark-baryon sector has been reported from an analysis of \mbox{$\Lb \to \proton \pim\pip\pim$} decays~\cite{LHCb-PAPER-2016-030}.
Other \CP-asymmetry parameters measured in $\Lb$ baryon decays to $\proton\pim$, $\proton\Km$~\cite{Aaltonen:2014vra}, $\KS\proton\pim$~\cite{LHCb-PAPER-2013-061}, $\Lz\Kp\Km$ and $\Lz\Kp\pim$~\cite{LHCb-PAPER-2016-004} final states are consistent with zero within the current experimental precision; these comprise the only charmless hadronic \bquark-baryon decays that have been observed to date.
It is therefore of great interest to search for additional charmless \bquark-baryon decays that may be used in future to investigate \CP-violation effects.

\begin{figure}[!b]
  \centering
  \includegraphics[width=0.40\textwidth]{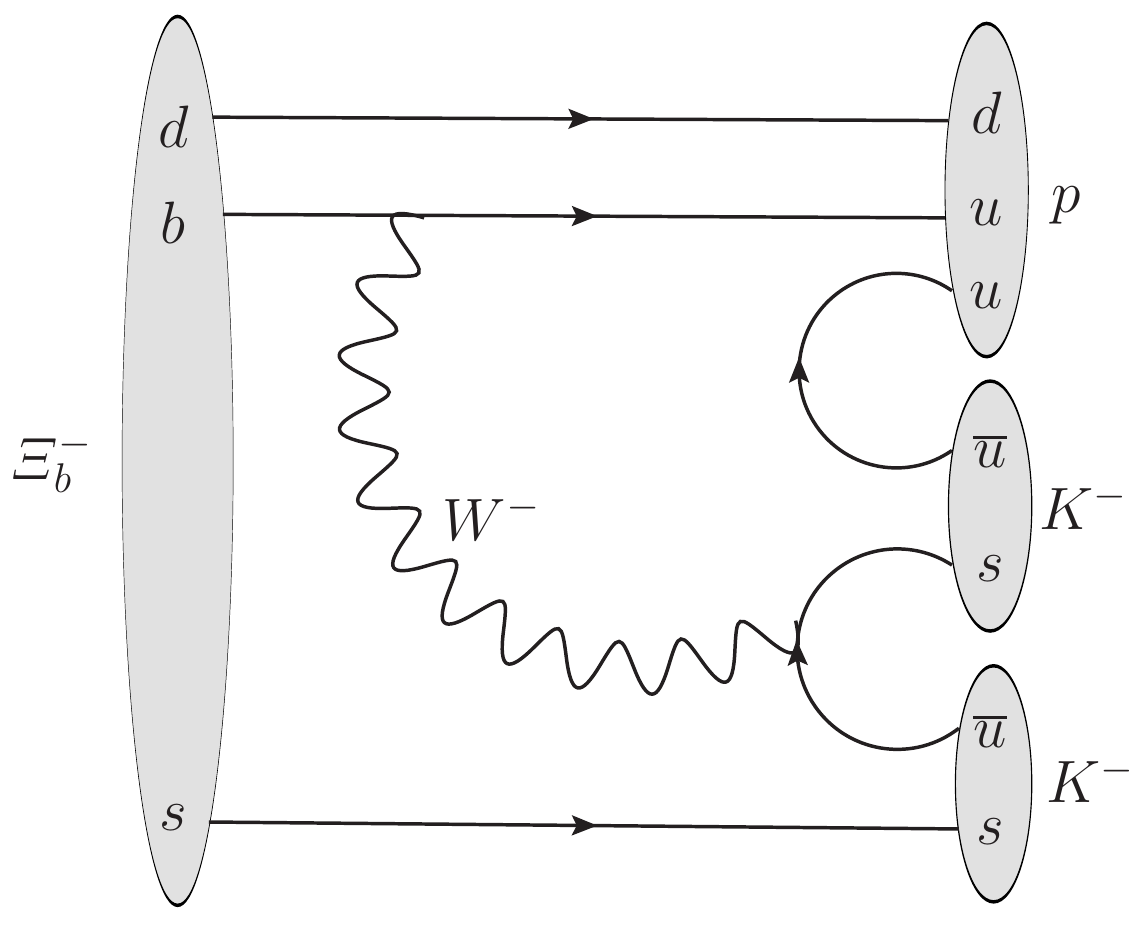}
  \includegraphics[width=0.40\textwidth]{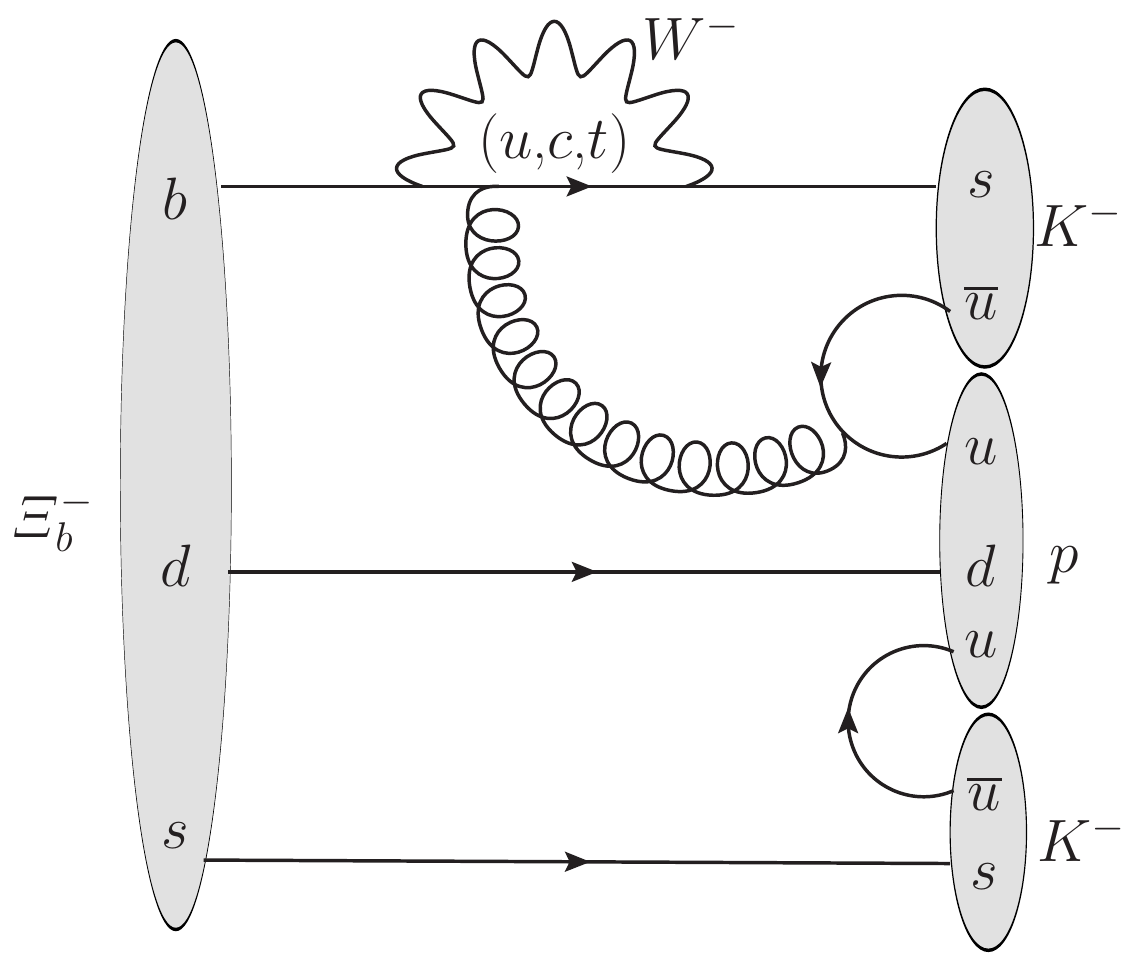}
  \caption{(Left) tree and (right) loop diagrams for the \mbox{$\Xibm \to \proton \Km\Km$} decay channel.} 
  \label{fig:feynman}
\end{figure}

In this Letter, the first search is presented for decays of $\Xibm$ and $\Omegab$ baryons, with constituent quark contents of $\bquark\squark\dquark$ and $\bquark\squark\squark$, to the charmless hadronic final states $\proton h^{-}h^{\prime -}$, where $h^{(\prime)}$ is a kaon or pion.
The inclusion of charge-conjugate processes is implied throughout.
Example decay diagrams for the \mbox{$\Xibm \to \proton \Km\Km$} mode are shown in Fig.~\ref{fig:feynman}.
Interference between Cabibbo-suppressed tree and loop diagrams may lead to \CP-violation effects.  
The \mbox{$\Xibm \to \proton \Km\pim$} and \mbox{$\Omegab \to \proton \Km\Km$} decays proceed by tree-level diagrams similar to that of Fig.~\ref{fig:feynman}~(left).
Diagrams for \mbox{$\Omegab \to \proton \Km\pim$} and both $\Xibm$ and \mbox{$\Omegab \to \proton \pim\pim$} require additional weak interaction vertices. The rates of these decays are therefore expected to be further suppressed.

The analysis is based on a sample of proton-proton collision data, recorded by the LHCb experiment at centre-of-mass energies $\sqrt{s} = 7$ and $8 \tev$, corresponding to $3 \invfb$ of integrated luminosity.
Since the fragmentation fractions, $f_\Xibm$ and $f_\Omegab$, which quantify the probabilities for a \bquark quark to hadronise into these particular states, have not been determined, it is not possible to measure absolute branching fractions.
Instead, the product of each branching fraction and the relevant fragmentation fraction is determined relative to the corresponding values for the topologically similar normalisation channel \mbox{$\Bm\to\Kp\Km\Km$} (the \Bm\ fragmentation fraction is denoted $f_u$).
Once one significant signal yield is observed, it becomes possible to determine ratios of branching fractions for decays of the same baryon to different final states, thus cancelling the dependence on the fragmentation fraction.

The LHCb detector~\cite{Alves:2008zz,LHCb-DP-2014-002} is a single-arm forward spectrometer covering the pseudorapidity range $2 < \eta < 5$, designed for the study of particles containing \bquark\ or \cquark\ quarks. 
The pseudorapidity is defined as $-\ln \left[ \tan(\theta/2) \right]$ where $\theta$ is the polar angle relative to the beam axis.
The detector elements that are particularly relevant to this analysis are: a silicon-strip vertex detector surrounding the $pp$ interaction region that allows \bquark\ hadrons to be identified from their characteristically long flight distance; a tracking system that provides a measurement of the momentum ($p$) of charged particles; two ring-imaging Cherenkov detectors that enable different species of charged hadrons to be distinguished; and calorimeter and muon systems that provide information used for online event selection.
Simulated data samples, produced with software described in Refs.~\cite{Sjostrand:2006za,*Sjostrand:2007gs,LHCb-PROC-2010-056,Lange:2001uf,Golonka:2005pn,Allison:2006ve,*Agostinelli:2002hh,LHCb-PROC-2011-006}, are used to evaluate the response of the detector to signal decays and to characterise the properties of certain types of background. 
These samples are generated separately for centre-of-mass energies of $7$ and $8 \tev$, simulating the corresponding data-taking conditions, and combined in appropriate quantities.

Online event selection is performed by a trigger~\cite{LHCb-DP-2012-004} that consists of a hardware stage, based on information from the calorimeter and muon systems, followed by a software stage, which applies a full event reconstruction.
At the hardware trigger stage, events are required to contain either a muon with high transverse momentum (\pt) or a particle that deposits high transverse energy in the calorimeters.
For hadrons, the transverse energy threshold is typically $3.5 \gev$. 
The software trigger for this analysis requires a two- or three-track secondary vertex with significant displacement from the primary $pp$ interaction vertices (PVs). 
At least one charged particle must have \pt\ above a threshold of $1.7\,(1.6) \gevc$ in the $\sqrt{s} = 7\,(8) \tev$ data. 
This particle must also be inconsistent with originating from any PV as quantified through the difference in the vertex-fit \chisq of a given PV reconstructed with and without the considered particle (\chisqip).
A multivariate algorithm~\cite{BBDT} is used for the identification of secondary vertices consistent with the decay of a \bquark hadron.

The offline selection of \bquark-hadron candidates formed from three tracks is carried out with an initial prefiltering stage, a requirement on the output of a neural network~\cite{Feindt2006190}, and particle identification criteria.
To avoid potential bias, the properties of candidates with invariant masses in windows around the $\Xibm$ and $\Omegab$ masses were not inspected until after the analysis procedures were finalised.
The prefiltering includes requirements on the quality, $p$, \pt\ and \chisqip\ of the tracks.
Each \bquark candidate must have a good quality vertex that is displaced from the closest PV (\ie that with which it forms the smallest \chisqip), must satisfy $p$ and \pt\ requirements, and must have reconstructed invariant mass loosely consistent with those of the \bquark hadrons.
A requirement is also imposed on the angle $\theta_{\rm dir}$ between the \bquark-candidate momentum vector and the line between the PV and the \bquark-candidate decay vertex.
In the offline selection, trigger signals are associated with reconstructed particles.
Selection requirements can therefore be made not only on which trigger caused the event to be recorded, but also on whether the decision was due to the signal candidate or other particles produced in the $pp$ collision~\cite{LHCb-DP-2012-004}.
Only candidates from events with a hardware trigger caused by deposits of the signal in the calorimeter, or caused by other particles in the event, are retained.
It is also required that the software trigger decision must have been caused by the signal candidate.

The inputs to the neural network for the final selection are the scalar sum of the \pt\ of all final-state tracks, the values of \pt\ and \chisqip\ for the highest \pt\ final-state track, the \bquark-candidate $\cos\left(\theta_{\rm dir}\right)$, vertex \chisq and \chisqip, together with a combination of momentum information and $\theta_{\rm dir}$ that characterises how closely the momentum vector of the \bquark candidate points back to the PV.
The $\pt$ asymmetry between the \bquark candidate and other tracks within a circle, centred on the \bquark candidate, with a radius $R = \sqrt{\delta\eta^2 + \delta\phi^2} < 1.5$ in the space of pseudorapidity and azimuthal angle $\phi$ (in radians) around the beam direction~\cite{LHCb-PAPER-2012-001} is also used in the network.
The distributions of these variables are consistent between simulated samples of signal decays and the \mbox{$\Bm\to\Kp\Km\Km$} normalisation channel, and between background-subtracted \mbox{$\Bm\to\Kp\Km\Km$} data and simulation.
The neural network input variables are also found to be not strongly correlated with either the \bquark-candidate mass or the position in the phase space of the decay.
The neural network is trained to distinguish signal from combinatorial background in the \mbox{$\Bm\to\Kp\Km\Km$} channel, using a data-driven approach in which the two components are separated statistically using the \sPlot\ method~\cite{Pivk:2004ty} with the \bquark-candidate mass as discriminating variable.
The requirement on the neural network output is optimised using a figure of merit~\cite{Punzi:2003bu} intended to give the best chance to observe the signal decays.
The same neural network output requirement is made for all signal final states, and has an efficiency of about $60\,\%$.

Using information from the ring-imaging Cherenkov detectors~\cite{LHCb-DP-2012-003}, criteria that identify uniquely the final-state tracks as either protons, pions or kaons are imposed, ensuring that no candidate appears in more than one of the final states considered. 
For pions and kaons these criteria are optimised simultaneously with that on the neural network output, using the same figure of merit.
The desire to reject possible background from \mbox{$\Bm \to \Kp h^{-}h^{\prime -}$} in the signal modes justifies independent treatment of the proton identification requirement.
In the simultaneous optimisation, the efficiency is taken from control samples while the expected background level is extrapolated from sidebands in the \bquark-candidate mass distribution.
The combined efficiency of the particle identification requirements is about $30\,\%$ for the $\proton\Km\Km$, $40\,\%$ for the $\proton\Km\pim$ and $50\,\%$ for the $\proton \pim\pim$ final state.

In order to ensure that any signal seen is due to charmless decays, candidates with $\proton \Km$ invariant mass consistent with the \mbox{$\Xibm \to \Xicz h^- \to \proton \Km h^-$} or \mbox{$\Xibm \to \Xicz h^- \to \proton \pim h^-$} decay chain are vetoed.
Similarly, candidates for the normalisation channel with $\Kp\Km$ invariant mass consistent with the \mbox{$\Bm \to \Dz\Km \to \Kp\Km\Km$} decay chain are removed.
After all selection requirements are imposed, the fraction of selected events that contain more than one candidate is much less than $1\,\%$; all such candidates are retained.

The yields of the signal decays are obtained from a simultaneous unbinned extended maximum likelihood fit to the \bquark-candidate mass distributions in the three $\proton h^- h^{\prime -}$ final states.
This approach allows potential cross-feed from one channel to another, due to particle misidentification, to be constrained according to the expected rates.
The yield of the normalisation channel is determined from a separate fit to the $\Kp\Km\Km$ mass distribution.

Each signal component is modelled with the sum of two Crystal Ball (CB) functions~\cite{Skwarnicki:1986xj} with shared parameters describing the core width and peak position and with non-Gaussian tails to both sides.
The tail parameters and the relative normalisation of the CB functions are determined from simulation.
A scale factor relating the width in data to that in simulation is determined from the fit to the normalisation channel.
In the fit to the signal modes the peak positions are fixed to the known \Xibm\ and \Omegab\ masses~\cite{LHCb-PAPER-2014-048,LHCb-PAPER-2016-008,HFAG}; the only free parameters associated with the signal components are the yields.

Cross-feed backgrounds from other decays to $\proton h^- h^{\prime -}$ final states are also modelled with the sum of two CB functions, with all shape parameters fixed according to simulation but the width scaled in the same way as signal components.
Cross-feed backgrounds from \mbox{$\Bm \to \Kp h^{-}h^{\prime -}$} decays are modelled, in the mass interval of the fit, by exponential functions with shape fixed according to simulation.
The yields of all cross-feed backgrounds are constrained according to expectations based on the yield in the correctly reconstructed channel and the (mis-)identification probabilities determined from control samples.

In addition to signal and cross-feed backgrounds, components for partially reconstructed and combinatorial backgrounds are included in each final state.
Partially reconstructed backgrounds arise due to \bquark-hadron decays into final states similar to the signal, but with additional soft particles that are not reconstructed.
Possible examples include \mbox{$\Xibm \to N^+ h^{-}h^{\prime -} \to \proton \piz h^{-}h^{\prime -}$} and \mbox{$\Xibm \to \proton \Kstarm h^- \to \proton \Km \piz h^-$}.
Such decays are investigated with simulation and it is found that many of them have similar \bquark-candidate mass distributions.
The shapes of these backgrounds are therefore taken from \mbox{$\Xibm \to N^+ h^{-}h^{\prime -} \to \proton \piz h^{-}h^{\prime -}$} simulation, with possible additional contributions considered as a source of systematic uncertainty.
The shapes are modelled with an ARGUS function~\cite{Albrecht:1990am} convolved with a Gaussian function.
The parameters of these functions are taken from simulation, except for the threshold of the ARGUS function, which is fixed to the known mass difference $m_{\Xibm}-m_{\piz}$~\cite{PDG2014,LHCb-PAPER-2014-048}.
The combinatorial background is modelled by an exponential function with the shape parameter shared between the three final states.
Possible differences in the shape between the different final states are considered as a source of systematic uncertainty.
The free parameters of the fit are the signal and background yields, and the combinatorial background shape parameter.
The stability of the fit is confirmed using ensembles of pseudoexperiments with different values of signal yields.

\begin{figure}[!tb]
  \centering
  \includegraphics[width=0.49\textwidth]{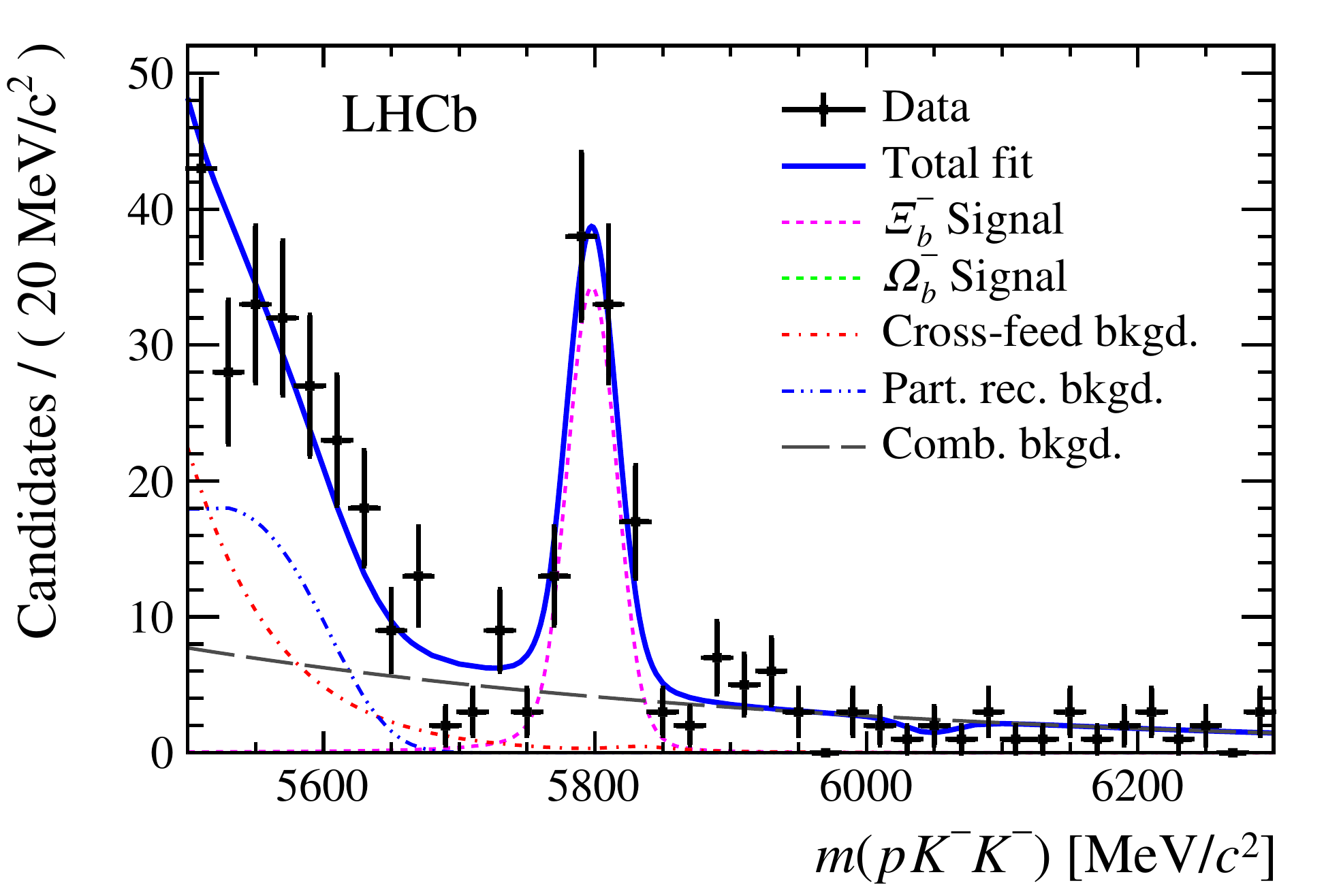}
  \includegraphics[width=0.49\textwidth]{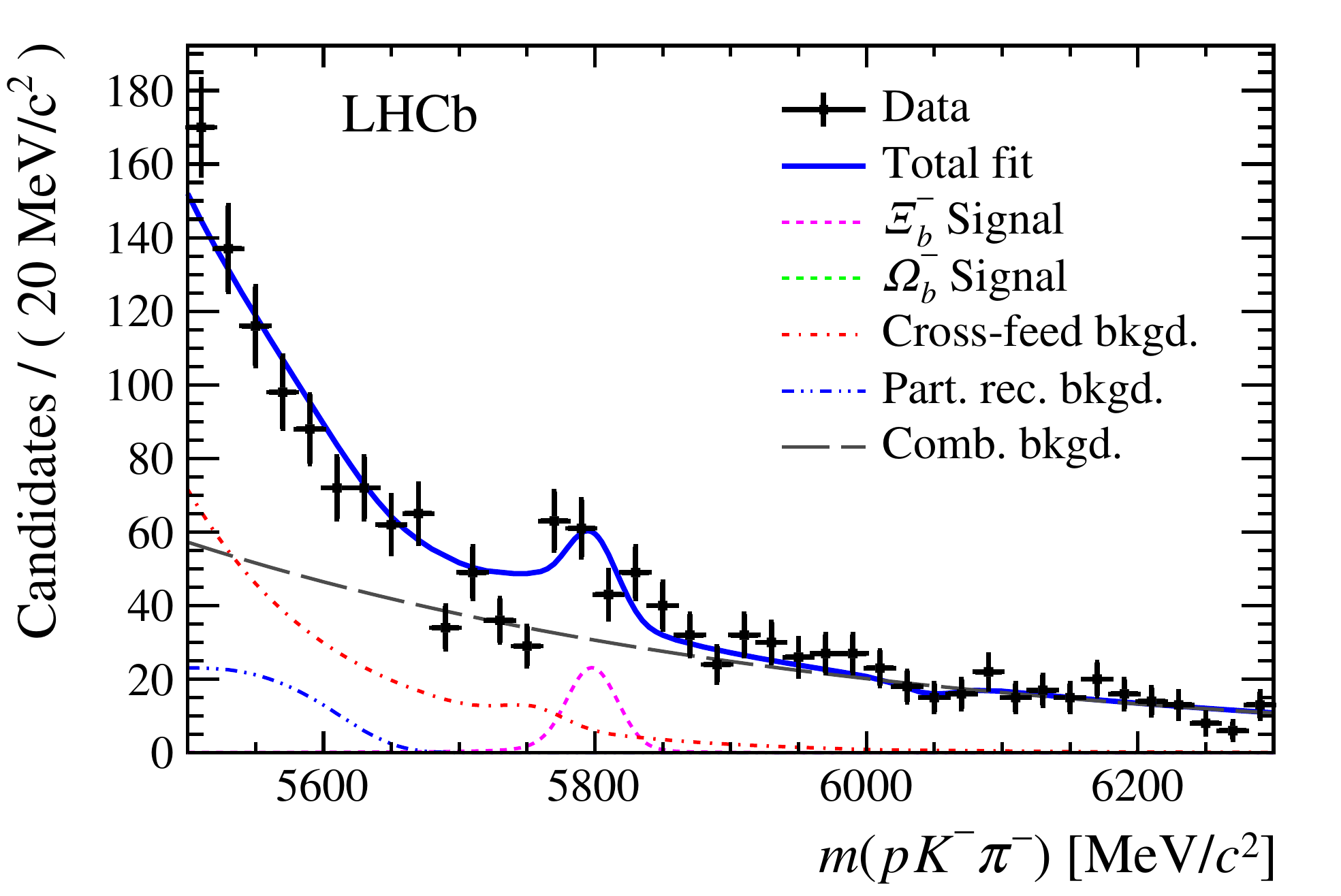}
  \includegraphics[width=0.49\textwidth]{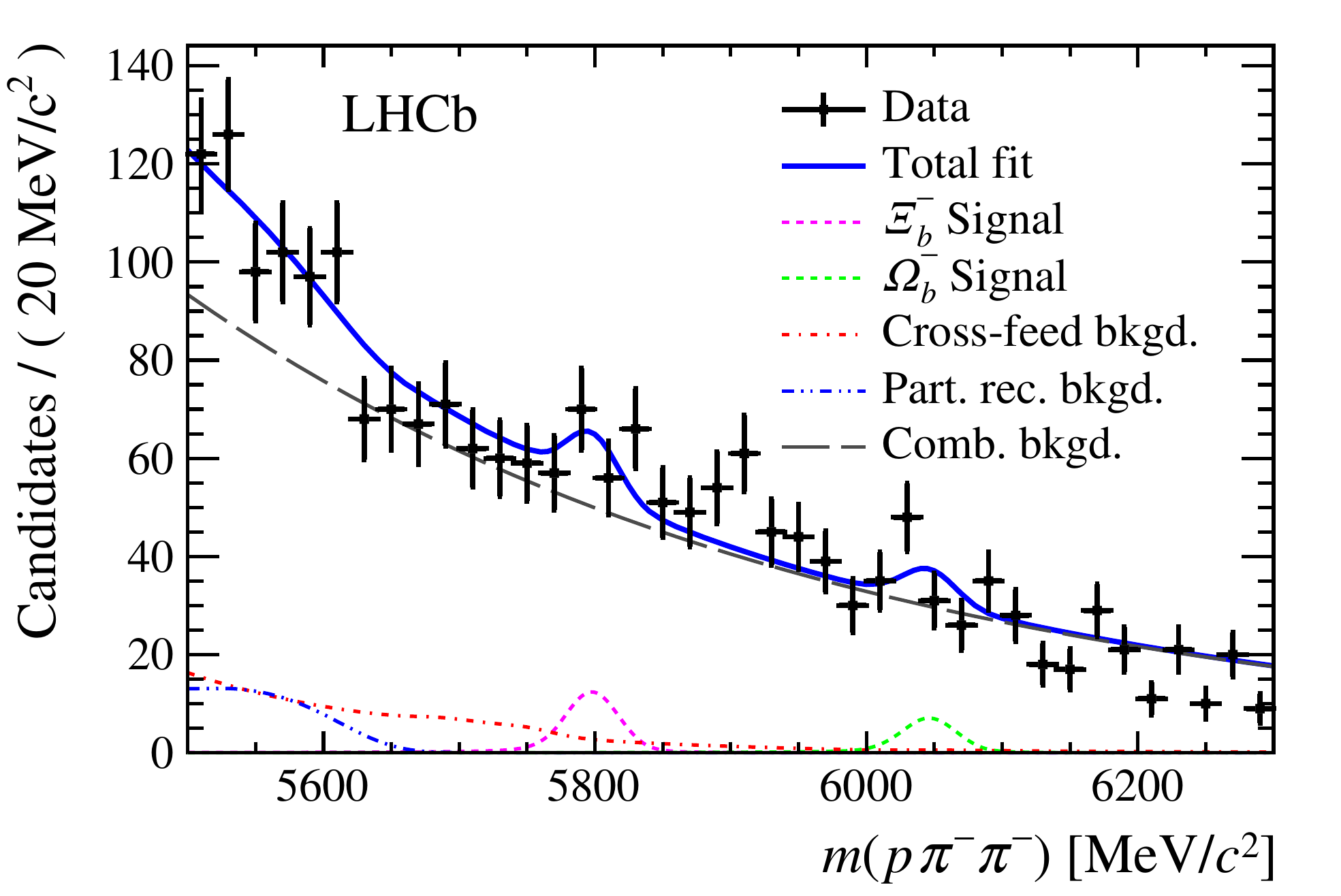}
  \includegraphics[width=0.49\textwidth]{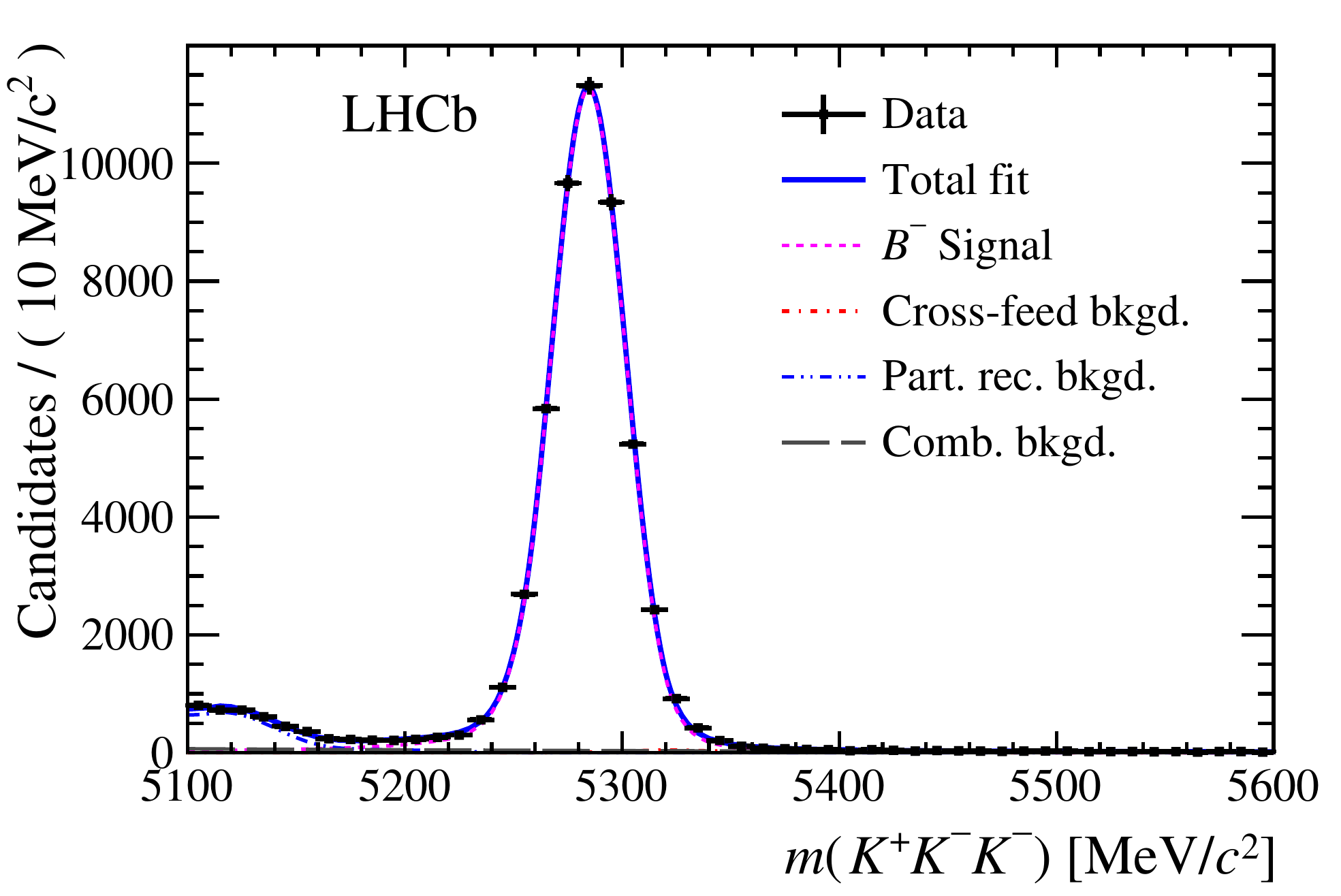}
  \caption{
    Mass distributions for \bquark-hadron candidates in the (top left) $\proton \Km\Km$, (top right) $\proton \Km\pim$, (bottom left) $\proton \pim\pim$ and (bottom right) $\Kp\Km\Km$ final states.
    Results of the fits are shown with dark blue solid lines.
    Signals for $\Xibm$ and $\Bm$ ($\Omegab$) decays are shown with pink (light green) dashed lines, combinatorial backgrounds are shown with grey long-dashed lines, cross-feed backgrounds are shown with red dot-dashed lines, and partially reconstructed backgrounds are shown with dark blue double-dot-dashed lines.
  } 
  \label{fig:fits}
\end{figure}

The results of the fits are shown in Fig.~\ref{fig:fits}.
The significance of each of the signals is determined from the change in likelihood when the corresponding yield is fixed to zero, with relevant sources of systematic uncertainty taken into account.
The signals for $\Xibm \to \proton \Km\Km$ and $\proton \Km\pim$ decays are found to have significance of $8.7\,\sigma$ and $3.4\,\sigma$, respectively; each of the other signal modes has significance less than $2\,\sigma$.
The relative branching fractions multiplied by fragmentation fractions are determined as
\begin{equation}
  R_{\proton h^- h^{\prime -}} \equiv \frac{f_{\Xibm}}{f_{u}} \frac{\BF(\Xibm \to \proton h^- h^{\prime -})}{\BF(\Bm \to \Kp\Km\Km)} = \frac{\mathcal{N}(\Xibm \to \proton h^- h^{\prime -})}{\mathcal{N}(\Bm\to \Kp\Km\Km)} \frac{\epsilon(\Bm\to \Kp\Km\Km)}{\epsilon(\Xibm \to \proton h^- h^{\prime -})} \, ,
  \label{bfeq}
\end{equation}
where the yields $\mathcal{N}$ are obtained from the fits.
A similar expression is used for the $\Omegab$ decay modes.
The efficiencies $\epsilon$ are determined from simulation, weighted according to the most recent $\Xibm$ and $\Omegab$ lifetime measurements~\cite{LHCb-PAPER-2014-048,LHCb-PAPER-2016-008,HFAG}, taking into account contributions from the detector geometry, reconstruction and both online and offline selection criteria. 
These are determined as a function of the position in phase space in each of the three-body final states.
The phase space for each of the \Xibm\ and \Omegab\ decays to $\proton h^- h^{\prime -}$ is five-dimensional, but significant variations in efficiency occur only in the variables that describe the Dalitz plot.
Simulation is used to evaluate each contribution to the efficiency except for the effect of the particle identification criteria, which is determined from data control samples weighted according to the expected kinematics of the signal tracks~\cite{LHCb-DP-2012-003,Anderlini:2202412}.
The description of reconstruction and selection efficiencies in the simulation has been validated with large control samples; the impact on the results of possible residual differences between data and simulation is negligible.  

\begin{figure}[!tb]
  \centering
  \includegraphics[width=0.6\textwidth]{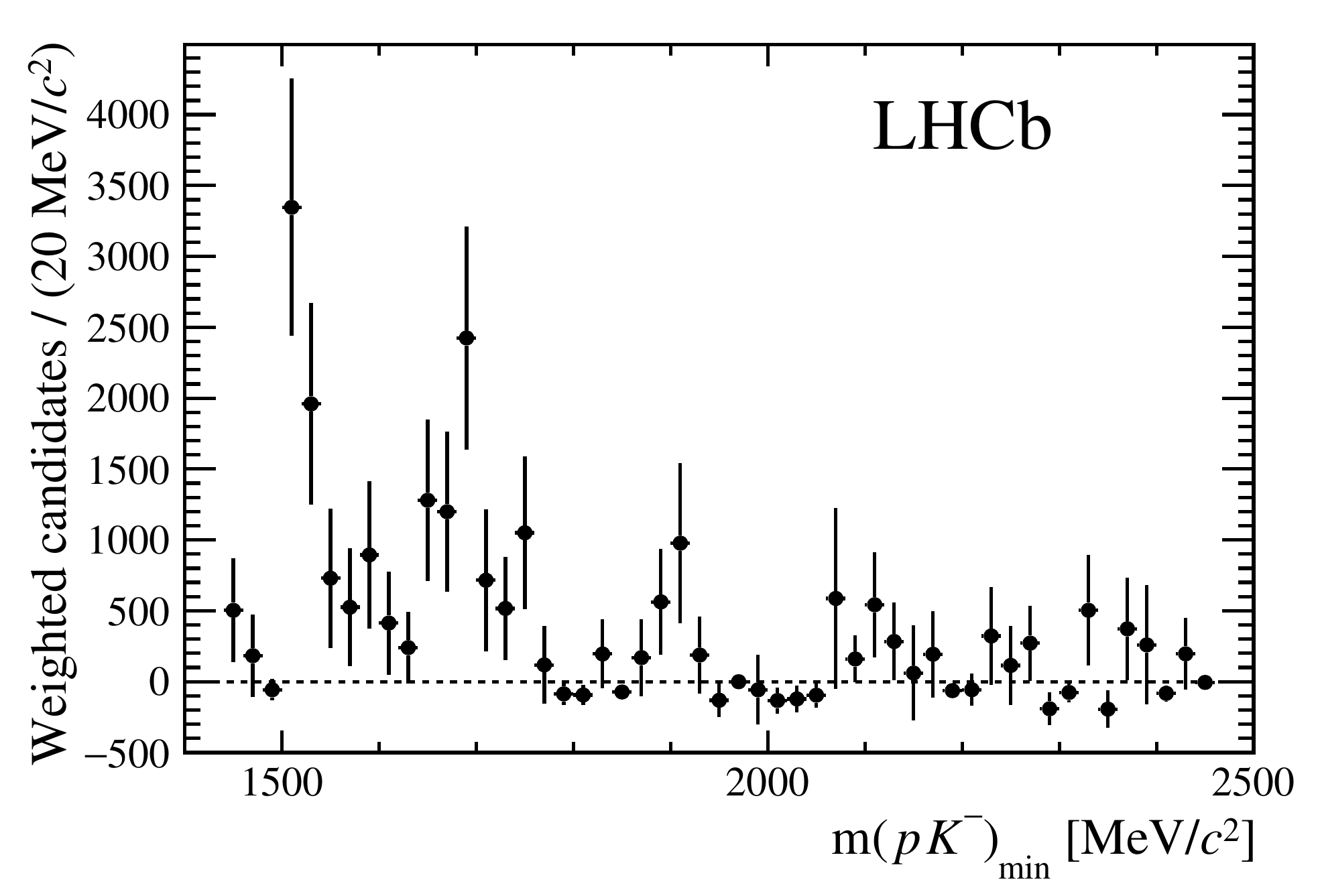}
  \caption{
    Efficiency-corrected and background-subtracted~\cite{Pivk:2004ty} $m(\proton\Km)_{\rm min}$ distribution from \mbox{$\Xibm \to \proton\Km\Km$} candidates.
  }
  \label{fig:mpK}
\end{figure}

For the \mbox{$\Xibm \to \proton\Km\Km$}, \mbox{$\Xibm \to \proton\Km\pim$} and \mbox{$\Bm\to \Kp\Km\Km$} channels, efficiency corrections for each candidate are applied using the method of Ref.~\cite{LHCb-PAPER-2012-018} to take the variation over the phase space into account.
Using this procedure, the efficiency-corrected and background-subtracted $m(\proton\Km)_{\rm min}$ distribution shown in Fig.~\ref{fig:mpK} is obtained from \mbox{$\Xibm \to \proton\Km\Km$} candidates.
Here $m(\proton\Km)_{\rm min}$ indicates the smaller of the two $m(\proton\Km)$ values for each signal candidate, evaluated with the $\Xibm$ and the final-state particle masses fixed to their known values~\cite{PDG2014,LHCb-PAPER-2014-048}.
The distribution contains a clear peak from the $\Lz(1520)$ resonance, a structure that is consistent with being a combination of the $\Lz(1670)$ and $\Lz(1690)$ states, and possible additional contributions at higher mass.
Compared to the $\proton \Km$ structures seen in the amplitude analysis of $\Lb \to \jpsi \proton \Km$~\cite{LHCb-PAPER-2015-029}, the contributions from the broad $\Lz(1600)$ and $\Lz(1810)$ states appear to be smaller.
A detailed amplitude analysis will be of interest when larger samples are available.

For channels without significant signal yields the efficiency averaged over phase space is used in Eq.~(\ref{bfeq}).
A corresponding systematic uncertainty is assigned from the variation of the efficiency over the phase space; this is the dominant source of systematic uncertainty for those channels.
The quantities entering Eq.~(\ref{bfeq}), and the results for $R_{\proton h^- h^{\prime -}}$, are reported in Table~\ref{tab:results}.
When the signal significance is less than $3\,\sigma$, upper limits are set by integrating the likelihood after multiplying by a prior probability distribution that is uniform in the region of positive branching fraction.

\begin{table}[!tb]
  \centering
  \caption{
    Fitted yields, efficiencies and relative branching fractions multiplied by fragmentation fractions ($R_{\proton h^- h^{\prime -}}$).
    The two uncertainties quoted on $R_{\proton h^- h^{\prime -}}$ are statistical and systematic.
    Upper limits are quoted at 90~(95)\,\% confidence level for modes with signal significance less than $3\,\sigma$.
    Uncertainties on the efficiencies are not given as only the relative uncertainties affect the branching fraction measurements.
  }
  \label{tab:results}
\resizebox{\textwidth}{!}{
  \begin{tabular}{lr@{$\;\pm\;$}lcr@{$\;\pm\;$}l@{$\;\pm\;$}lc}
    \hline \\ [-2.7ex]
    Mode & \multicolumn{2}{c}{Yield ${\cal N}$} & Efficiency $\epsilon$ (\%) & \multicolumn{4}{c}{$R_{\proton h^- h^{\prime -}} \ (10^{-5})$} \\
    \hline
    $\Xibm \to \proton \Km\Km$  &  82.9  & 10.4 & 0.398 & $\phantom{\!-}265$ & 35 & 47 \\
    $\Xibm \to \proton \Km\pim$ &  59.6  & 16.0 & 0.293 & $\phantom{\!-}259$ & 64 & 49 \\
    $\Xibm \to \proton\pim\pim$ &  33.2  & 17.9 & 0.573 & $\phantom{\!-1}74$ & 40 & 36 & $<$ 147 ~(166)\\
    $\Omegab\to\proton\Km\Km$   & $-2.8$ & \phantom{1}2.5 & 0.375 & $\phantom{\!1}-9$ & \phantom{0}9 & \phantom{0}6 & $<$ \phantom{1}18 ~\phantom{1}(22)\\
    $\Omegab\to\proton\Km\pim$  & $-7.6$ & \phantom{1}9.2 & 0.418 & $\phantom{\!1}{-23}$ & 28 & 23 & $<$ \phantom{1}51 ~\phantom{1}(62)\\
    $\Omegab\to\proton\pim\pim$ &  20.1  & 13.8 & 0.536 & $\phantom{\!-0}48$ & 33 & 28 & $<$ 109 ~(124) \\
    $\Bm \to \Kp\Km\Km$        & 50\,490 &  250 & 0.643 & \multicolumn{3}{c}{---}\\
    \hline
  \end{tabular}
}
\end{table}

The sources of systematic uncertainty arise from the fit model and the knowledge of the efficiency.
The fit model is changed by varying the fixed parameters of the model, using alternative shapes for the components, and by including components that are omitted in the baseline fit.  
Intrinsic biases in the fitted yields are investigated with simulated pseudoexperiments, and are found to be negligible.
Uncertainties in the efficiency arise due to the limited size of the simulation samples and possible residual differences between data and simulation in the trigger and particle identification efficiencies~\cite{LHCb-PAPER-2014-036}.
Possible biases in the results due to the vetoes of charm hadrons are also accounted for.
The efficiency depends on the signal decay-time distribution, and therefore the precision of the \Xibm\ and \Omegab\ lifetime measurements~\cite{LHCb-PAPER-2014-048,LHCb-PAPER-2016-008,HFAG} is a source of uncertainty.
Similarly, the \pt\ distribution assumed for signal decays in the simulation affects the efficiency.
Since the \pt\ spectra for \Xibm\ and \Omegab\ baryons produced in LHC collisions have not been measured, the effect is estimated by weighting simulation to the background-subtracted~\cite{Pivk:2004ty} \pt\ distribution for \mbox{$\Xibm \to \proton\Km\Km$} decays obtained from the data.
The difference in the average efficiency between the weighted and unweighted simulation is assigned as the associated systematic uncertainty.
This is the dominant source of systematic uncertainty for the \mbox{$\Xibm \to \proton\Km\Km$} and \mbox{$\Xibm \to \proton\Km\pim$} modes.

The yield of $\Xibm \to \proton \Km\Km$ decays is sufficient to use as normalisation for the relative branching fractions of the other $\Xibm$ decays.
The results are
\begin{eqnarray*}
  \frac{{\cal B}(\Xibm \to \proton \Km\pim)}{{\cal B}(\Xibm \to \proton \Km\Km) } & = & 0.98 \pm 0.27 \stat \pm 0.09 \syst \,, \\
  \frac{{\cal B}(\Xibm \to \proton \pim\pim)}{{\cal B}(\Xibm \to \proton \Km\Km) } & = & 0.28 \pm 0.16 \stat \pm 0.13 \syst \ \ \ < \ 0.56~(0.63) \, ,
\end{eqnarray*}
where the upper limit is quoted at 90~(95)\,\% confidence level.
The same sources of systematic uncertainty as discussed above are considered.
Since the effects due to the \pt\ distribution largely cancel, the dominant contributions are due to the trigger efficiency, fit model and (for the $\Xibm \to \proton \pim\pim$ mode) efficiency variation across the phase space.

In summary, a search for decays of $\Xibm$ and $\Omegab$ baryons to $\proton h^{-}h^{\prime -}$ final states has been carried out with a sample of proton-proton collision data corresponding to an integrated luminosity of $3 \invfb$.
The first observation of the \mbox{$\Xibm \to \proton \Km\Km$} decay, and first evidence for the \mbox{$\Xibm \to \proton \Km\pim$} decay, have been obtained; there is no significant signal for the other modes.
This is the first observation of a $\Xib$ decay to a charmless final state.
These modes may be used in future to search for \CP asymmetries in the \bquark-baryon sector.

\section*{Acknowledgements}

\noindent We express our gratitude to our colleagues in the CERN
accelerator departments for the excellent performance of the LHC. We
thank the technical and administrative staff at the LHCb
institutes. We acknowledge support from CERN and from the national
agencies: CAPES, CNPq, FAPERJ and FINEP (Brazil); NSFC (China);
CNRS/IN2P3 (France); BMBF, DFG and MPG (Germany); INFN (Italy); 
FOM and NWO (The Netherlands); MNiSW and NCN (Poland); MEN/IFA (Romania); 
MinES and FASO (Russia); MinECo (Spain); SNSF and SER (Switzerland); 
NASU (Ukraine); STFC (United Kingdom); NSF (USA).
We acknowledge the computing resources that are provided by CERN, IN2P3 (France), KIT and DESY (Germany), INFN (Italy), SURF (The Netherlands), PIC (Spain), GridPP (United Kingdom), RRCKI and Yandex LLC (Russia), CSCS (Switzerland), IFIN-HH (Romania), CBPF (Brazil), PL-GRID (Poland) and OSC (USA). We are indebted to the communities behind the multiple open 
source software packages on which we depend.
Individual groups or members have received support from AvH Foundation (Germany),
EPLANET, Marie Sk\l{}odowska-Curie Actions and ERC (European Union), 
Conseil G\'{e}n\'{e}ral de Haute-Savoie, Labex ENIGMASS and OCEVU, 
R\'{e}gion Auvergne (France), RFBR and Yandex LLC (Russia), GVA, XuntaGal and GENCAT (Spain), Herchel Smith Fund, The Royal Society, Royal Commission for the Exhibition of 1851 and the Leverhulme Trust (United Kingdom).

\addcontentsline{toc}{section}{References}
\setboolean{inbibliography}{true}
\bibliographystyle{LHCb}

\begin{mcitethebibliography}{10}
\mciteSetBstSublistMode{n}
\mciteSetBstMaxWidthForm{subitem}{\alph{mcitesubitemcount})}
\mciteSetBstSublistLabelBeginEnd{\mcitemaxwidthsubitemform\space}
{\relax}{\relax}

\bibitem{Lees:2012mma}
BaBar collaboration, J.~P. Lees {\em et~al.},
  \ifthenelse{\boolean{articletitles}}{\emph{{Measurement of \CP asymmetries
  and branching fractions in charmless two-body $B$-meson decays to pions and
  kaons}}, }{}\href{http://dx.doi.org/10.1103/PhysRevD.87.052009}{Phys.\ Rev.\
  \textbf{D87} (2013) 052009},
  \href{http://arxiv.org/abs/1206.3525}{{\normalfont\ttfamily
  arXiv:1206.3525}}\relax
\mciteBstWouldAddEndPuncttrue
\mciteSetBstMidEndSepPunct{\mcitedefaultmidpunct}
{\mcitedefaultendpunct}{\mcitedefaultseppunct}\relax
\EndOfBibitem
\bibitem{Duh:2012ie}
Belle collaboration, Y.-T. Duh {\em et~al.},
  \ifthenelse{\boolean{articletitles}}{\emph{{Measurements of branching
  fractions and direct \CP asymmetries for $B\to K\pi$, $B\to \pi\pi$ and $B
  \to KK$ decays}},
  }{}\href{http://dx.doi.org/10.1103/PhysRevD.87.031103}{Phys.\ Rev.\
  \textbf{D87} (2013) 031103},
  \href{http://arxiv.org/abs/1210.1348}{{\normalfont\ttfamily
  arXiv:1210.1348}}\relax
\mciteBstWouldAddEndPuncttrue
\mciteSetBstMidEndSepPunct{\mcitedefaultmidpunct}
{\mcitedefaultendpunct}{\mcitedefaultseppunct}\relax
\EndOfBibitem
\bibitem{Aaltonen:2014vra}
CDF collaboration, T.~A. Aaltonen {\em et~al.},
  \ifthenelse{\boolean{articletitles}}{\emph{{Measurements of direct
  \CP-violating asymmetries in charmless decays of bottom baryons}},
  }{}\href{http://dx.doi.org/10.1103/PhysRevLett.113.242001}{Phys.\ Rev.\
  Lett.\  \textbf{113} (2014) 242001},
  \href{http://arxiv.org/abs/1403.5586}{{\normalfont\ttfamily
  arXiv:1403.5586}}\relax
\mciteBstWouldAddEndPuncttrue
\mciteSetBstMidEndSepPunct{\mcitedefaultmidpunct}
{\mcitedefaultendpunct}{\mcitedefaultseppunct}\relax
\EndOfBibitem
\bibitem{LHCb-PAPER-2013-018}
LHCb collaboration, R.~Aaij {\em et~al.},
  \ifthenelse{\boolean{articletitles}}{\emph{{First observation of $\CP$
  violation in the decays of $\Bs$ mesons}},
  }{}\href{http://dx.doi.org/10.1103/PhysRevLett.110.221601}{Phys.\ Rev.\
  Lett.\  \textbf{110} (2013) 221601},
  \href{http://arxiv.org/abs/1304.6173}{{\normalfont\ttfamily
  arXiv:1304.6173}}\relax
\mciteBstWouldAddEndPuncttrue
\mciteSetBstMidEndSepPunct{\mcitedefaultmidpunct}
{\mcitedefaultendpunct}{\mcitedefaultseppunct}\relax
\EndOfBibitem
\bibitem{LHCb-PAPER-2013-027}
LHCb collaboration, R.~Aaij {\em et~al.},
  \ifthenelse{\boolean{articletitles}}{\emph{{Measurement of $\CP$ violation in
  the phase space of $\Bpm\to\Kpm\pip\pim$ and $\Bpm\to\Kpm\Kp\Km$ decays}},
  }{}\href{http://dx.doi.org/10.1103/PhysRevLett.111.101801}{Phys.\ Rev.\
  Lett.\  \textbf{111} (2013) 101801},
  \href{http://arxiv.org/abs/1306.1246}{{\normalfont\ttfamily
  arXiv:1306.1246}}\relax
\mciteBstWouldAddEndPuncttrue
\mciteSetBstMidEndSepPunct{\mcitedefaultmidpunct}
{\mcitedefaultendpunct}{\mcitedefaultseppunct}\relax
\EndOfBibitem
\bibitem{LHCb-PAPER-2013-051}
LHCb collaboration, R.~Aaij {\em et~al.},
  \ifthenelse{\boolean{articletitles}}{\emph{{Measurement of $\CP$ violation in
  the phase space of $\Bpm\to\Kp\Km\pipm$ and $\Bpm\to\pip\pim\pipm$ decays}},
  }{}\href{http://dx.doi.org/10.1103/PhysRevLett.112.011801}{Phys.\ Rev.\
  Lett.\  \textbf{112} (2014) 011801},
  \href{http://arxiv.org/abs/1310.4740}{{\normalfont\ttfamily
  arXiv:1310.4740}}\relax
\mciteBstWouldAddEndPuncttrue
\mciteSetBstMidEndSepPunct{\mcitedefaultmidpunct}
{\mcitedefaultendpunct}{\mcitedefaultseppunct}\relax
\EndOfBibitem
\bibitem{LHCb-PAPER-2014-044}
LHCb collaboration, R.~Aaij {\em et~al.},
  \ifthenelse{\boolean{articletitles}}{\emph{{Measurement of $\CP$ violation in
  the three-body phase space of charmless $\Bpm$ decays}},
  }{}\href{http://dx.doi.org/10.1103/PhysRevD.90.112004}{Phys.\ Rev.\
  \textbf{D90} (2014) 112004},
  \href{http://arxiv.org/abs/1408.5373}{{\normalfont\ttfamily
  arXiv:1408.5373}}\relax
\mciteBstWouldAddEndPuncttrue
\mciteSetBstMidEndSepPunct{\mcitedefaultmidpunct}
{\mcitedefaultendpunct}{\mcitedefaultseppunct}\relax
\EndOfBibitem
\bibitem{Gronau:2008gu}
M.~Gronau and J.~L. Rosner,
  \ifthenelse{\boolean{articletitles}}{\emph{{Implications for \CP asymmetries
  of improved data on $\B \to \Kz\piz$}},
  }{}\href{http://dx.doi.org/10.1016/j.physletb.2008.08.004}{Phys.\ Lett.\
  \textbf{B666} (2008) 467},
  \href{http://arxiv.org/abs/0807.3080}{{\normalfont\ttfamily
  arXiv:0807.3080}}\relax
\mciteBstWouldAddEndPuncttrue
\mciteSetBstMidEndSepPunct{\mcitedefaultmidpunct}
{\mcitedefaultendpunct}{\mcitedefaultseppunct}\relax
\EndOfBibitem
\bibitem{Baek:2009hv}
S.~Baek {\em et~al.}, \ifthenelse{\boolean{articletitles}}{\emph{{Diagnostic
  for new physics in $B \to \pi K$ decays}},
  }{}\href{http://dx.doi.org/10.1016/j.physletb.2009.06.018}{Phys.\ Lett.\
  \textbf{B678} (2009) 97},
  \href{http://arxiv.org/abs/0905.1495}{{\normalfont\ttfamily
  arXiv:0905.1495}}\relax
\mciteBstWouldAddEndPuncttrue
\mciteSetBstMidEndSepPunct{\mcitedefaultmidpunct}
{\mcitedefaultendpunct}{\mcitedefaultseppunct}\relax
\EndOfBibitem
\bibitem{Ciuchini:2006kv}
M.~Ciuchini, M.~Pierini, and L.~Silvestrini,
  \ifthenelse{\boolean{articletitles}}{\emph{{New bounds on the CKM matrix from
  $B \to K \pi \pi$ Dalitz plot analyses}},
  }{}\href{http://dx.doi.org/10.1103/PhysRevD.74.051301}{Phys.\ Rev.\
  \textbf{D74} (2006) 051301},
  \href{http://arxiv.org/abs/hep-ph/0601233}{{\normalfont\ttfamily
  arXiv:hep-ph/0601233}}\relax
\mciteBstWouldAddEndPuncttrue
\mciteSetBstMidEndSepPunct{\mcitedefaultmidpunct}
{\mcitedefaultendpunct}{\mcitedefaultseppunct}\relax
\EndOfBibitem
\bibitem{Ciuchini:2006st}
M.~Ciuchini, M.~Pierini, and L.~Silvestrini,
  \ifthenelse{\boolean{articletitles}}{\emph{{Hunting the CKM weak phase with
  time-integrated Dalitz analyses of $\Bs \to K K \pi$ and $\Bs \to K \pi \pi$
  decays}}, }{}\href{http://dx.doi.org/10.1016/j.physletb.2006.12.043}{Phys.\
  Lett.\  \textbf{B645} (2007) 201},
  \href{http://arxiv.org/abs/hep-ph/0602207}{{\normalfont\ttfamily
  arXiv:hep-ph/0602207}}\relax
\mciteBstWouldAddEndPuncttrue
\mciteSetBstMidEndSepPunct{\mcitedefaultmidpunct}
{\mcitedefaultendpunct}{\mcitedefaultseppunct}\relax
\EndOfBibitem
\bibitem{Gronau:2006qn}
M.~Gronau, D.~Pirjol, A.~Soni, and J.~Zupan,
  \ifthenelse{\boolean{articletitles}}{\emph{{Improved method for CKM
  constraints in charmless three-body \B and \Bs decays}},
  }{}\href{http://dx.doi.org/10.1103/PhysRevD.75.014002}{Phys.\ Rev.\
  \textbf{D75} (2007) 014002},
  \href{http://arxiv.org/abs/hep-ph/0608243}{{\normalfont\ttfamily
  arXiv:hep-ph/0608243}}\relax
\mciteBstWouldAddEndPuncttrue
\mciteSetBstMidEndSepPunct{\mcitedefaultmidpunct}
{\mcitedefaultendpunct}{\mcitedefaultseppunct}\relax
\EndOfBibitem
\bibitem{Gronau:2007vr}
M.~Gronau, D.~Pirjol, A.~Soni, and J.~Zupan,
  \ifthenelse{\boolean{articletitles}}{\emph{{Constraint on $\bar{\rho},
  \bar{\eta}$ from $B \to K^* \pi$}},
  }{}\href{http://dx.doi.org/10.1103/PhysRevD.77.057504}{Phys.\ Rev.\
  \textbf{D77} (2008) 057504},
  \href{http://arxiv.org/abs/0712.3751}{{\normalfont\ttfamily
  arXiv:0712.3751}}\relax
\mciteBstWouldAddEndPuncttrue
\mciteSetBstMidEndSepPunct{\mcitedefaultmidpunct}
{\mcitedefaultendpunct}{\mcitedefaultseppunct}\relax
\EndOfBibitem
\bibitem{Bediaga:2006jk}
I.~Bediaga, G.~Guerrer, and J.~M. de~Miranda,
  \ifthenelse{\boolean{articletitles}}{\emph{{Extracting the quark mixing phase
  $\gamma$ from $\Bpm\to \Kpm \pip \pim$, $\Bz \to \KS \pip \pim$, and $\Bzb
  \to \KS \pip \pim$}},
  }{}\href{http://dx.doi.org/10.1103/PhysRevD.76.073011}{Phys.\ Rev.\
  \textbf{D76} (2007) 073011},
  \href{http://arxiv.org/abs/hep-ph/0608268}{{\normalfont\ttfamily
  arXiv:hep-ph/0608268}}\relax
\mciteBstWouldAddEndPuncttrue
\mciteSetBstMidEndSepPunct{\mcitedefaultmidpunct}
{\mcitedefaultendpunct}{\mcitedefaultseppunct}\relax
\EndOfBibitem
\bibitem{Gronau:2010dd}
M.~Gronau, D.~Pirjol, and J.~Zupan,
  \ifthenelse{\boolean{articletitles}}{\emph{{\CP asymmetries in $B \rightarrow
  K\pi, K^*\pi, \rho K$ decays}},
  }{}\href{http://dx.doi.org/10.1103/PhysRevD.81.094011}{Phys.\ Rev.\
  \textbf{D81} (2010) 094011},
  \href{http://arxiv.org/abs/1001.0702}{{\normalfont\ttfamily
  arXiv:1001.0702}}\relax
\mciteBstWouldAddEndPuncttrue
\mciteSetBstMidEndSepPunct{\mcitedefaultmidpunct}
{\mcitedefaultendpunct}{\mcitedefaultseppunct}\relax
\EndOfBibitem
\bibitem{Gronau:2010kq}
M.~Gronau, D.~Pirjol, and J.~L. Rosner,
  \ifthenelse{\boolean{articletitles}}{\emph{{Calculating phases between $B \to
  K^* \pi$ amplitudes}},
  }{}\href{http://dx.doi.org/10.1103/PhysRevD.81.094026}{Phys.\ Rev.\
  \textbf{D81} (2010) 094026},
  \href{http://arxiv.org/abs/1003.5090}{{\normalfont\ttfamily
  arXiv:1003.5090}}\relax
\mciteBstWouldAddEndPuncttrue
\mciteSetBstMidEndSepPunct{\mcitedefaultmidpunct}
{\mcitedefaultendpunct}{\mcitedefaultseppunct}\relax
\EndOfBibitem
\bibitem{Imbeault:2010xg}
M.~Imbeault, N.~R.-L. Lorier, and D.~London,
  \ifthenelse{\boolean{articletitles}}{\emph{{Measuring $\gamma$ in $B\to K \pi
  \pi$ decays}}, }{}\href{http://dx.doi.org/10.1103/PhysRevD.84.034041}{Phys.\
  Rev.\  \textbf{D84} (2011) 034041},
  \href{http://arxiv.org/abs/1011.4973}{{\normalfont\ttfamily
  arXiv:1011.4973}}\relax
\mciteBstWouldAddEndPuncttrue
\mciteSetBstMidEndSepPunct{\mcitedefaultmidpunct}
{\mcitedefaultendpunct}{\mcitedefaultseppunct}\relax
\EndOfBibitem
\bibitem{Bhattacharya:2015uua}
B.~Bhattacharya and D.~London,
  \ifthenelse{\boolean{articletitles}}{\emph{{Using U-spin to extract $\gamma$
  from charmless $B \to PPP$ decays}},
  }{}\href{http://dx.doi.org/10.1007/JHEP04(2015)154}{JHEP \textbf{04} (2015)
  154}, \href{http://arxiv.org/abs/1503.00737}{{\normalfont\ttfamily
  arXiv:1503.00737}}\relax
\mciteBstWouldAddEndPuncttrue
\mciteSetBstMidEndSepPunct{\mcitedefaultmidpunct}
{\mcitedefaultendpunct}{\mcitedefaultseppunct}\relax
\EndOfBibitem
\bibitem{Cabibbo:1963yz}
N.~Cabibbo, \ifthenelse{\boolean{articletitles}}{\emph{{Unitary symmetry and
  leptonic decays}},
  }{}\href{http://dx.doi.org/10.1103/PhysRevLett.10.531}{Phys.\ Rev.\ Lett.\
  \textbf{10} (1963) 531}\relax
\mciteBstWouldAddEndPuncttrue
\mciteSetBstMidEndSepPunct{\mcitedefaultmidpunct}
{\mcitedefaultendpunct}{\mcitedefaultseppunct}\relax
\EndOfBibitem
\bibitem{Kobayashi:1973fv}
M.~Kobayashi and T.~Maskawa, \ifthenelse{\boolean{articletitles}}{\emph{{\CP
  violation in the renormalizable theory of weak interaction}},
  }{}\href{http://dx.doi.org/10.1143/PTP.49.652}{Prog.\ Theor.\ Phys.\
  \textbf{49} (1973) 652}\relax
\mciteBstWouldAddEndPuncttrue
\mciteSetBstMidEndSepPunct{\mcitedefaultmidpunct}
{\mcitedefaultendpunct}{\mcitedefaultseppunct}\relax
\EndOfBibitem
\bibitem{LHCb-PAPER-2016-030}
LHCb collaboration, R.~Aaij {\em et~al.},
  \ifthenelse{\boolean{articletitles}}{\emph{{Probing matter-antimatter
  asymmetries in beauty baryon decays}},
  }{}\href{http://dx.doi.org/10.1038/nphys4021}{Nature Physics (2017) },
  \href{http://arxiv.org/abs/1609.05216}{{\normalfont\ttfamily
  arXiv:1609.05216}}\relax
\mciteBstWouldAddEndPuncttrue
\mciteSetBstMidEndSepPunct{\mcitedefaultmidpunct}
{\mcitedefaultendpunct}{\mcitedefaultseppunct}\relax
\EndOfBibitem
\bibitem{LHCb-PAPER-2013-061}
LHCb collaboration, R.~Aaij {\em et~al.},
  \ifthenelse{\boolean{articletitles}}{\emph{{Searches for $\Lb$ and $\Xibz$
  decays to $\KS\proton\pim$ and $\KS\proton\Km$ final states with first
  observation of the $\Lb \to\KS\proton\pim$ decay}},
  }{}\href{http://dx.doi.org/10.1007/JHEP04(2014)087}{JHEP \textbf{04} (2014)
  087}, \href{http://arxiv.org/abs/1402.0770}{{\normalfont\ttfamily
  arXiv:1402.0770}}\relax
\mciteBstWouldAddEndPuncttrue
\mciteSetBstMidEndSepPunct{\mcitedefaultmidpunct}
{\mcitedefaultendpunct}{\mcitedefaultseppunct}\relax
\EndOfBibitem
\bibitem{LHCb-PAPER-2016-004}
LHCb collaboration, R.~Aaij {\em et~al.},
  \ifthenelse{\boolean{articletitles}}{\emph{{Observations of
  $\Lb\to\Lz\Kp\pim$ and $\Lb\to\Lz\Kp\Km$ decays and searches for other $\Lb$
  and $\Xibz$ decays to $\Lz h^+h^-$ final states}},
  }{}\href{http://dx.doi.org/10.1007/JHEP05(2016)081}{JHEP \textbf{05} (2016)
  081}, \href{http://arxiv.org/abs/1603.00413}{{\normalfont\ttfamily
  arXiv:1603.00413}}\relax
\mciteBstWouldAddEndPuncttrue
\mciteSetBstMidEndSepPunct{\mcitedefaultmidpunct}
{\mcitedefaultendpunct}{\mcitedefaultseppunct}\relax
\EndOfBibitem
\bibitem{Alves:2008zz}
LHCb collaboration, A.~A. Alves~Jr.\ {\em et~al.},
  \ifthenelse{\boolean{articletitles}}{\emph{{The \lhcb detector at the LHC}},
  }{}\href{http://dx.doi.org/10.1088/1748-0221/3/08/S08005}{JINST \textbf{3}
  (2008) S08005}\relax
\mciteBstWouldAddEndPuncttrue
\mciteSetBstMidEndSepPunct{\mcitedefaultmidpunct}
{\mcitedefaultendpunct}{\mcitedefaultseppunct}\relax
\EndOfBibitem
\bibitem{LHCb-DP-2014-002}
LHCb collaboration, R.~Aaij {\em et~al.},
  \ifthenelse{\boolean{articletitles}}{\emph{{LHCb detector performance}},
  }{}\href{http://dx.doi.org/10.1142/S0217751X15300227}{Int.\ J.\ Mod.\ Phys.\
  \textbf{A30} (2015) 1530022},
  \href{http://arxiv.org/abs/1412.6352}{{\normalfont\ttfamily
  arXiv:1412.6352}}\relax
\mciteBstWouldAddEndPuncttrue
\mciteSetBstMidEndSepPunct{\mcitedefaultmidpunct}
{\mcitedefaultendpunct}{\mcitedefaultseppunct}\relax
\EndOfBibitem
\bibitem{Sjostrand:2006za}
T.~Sj\"{o}strand, S.~Mrenna, and P.~Skands,
  \ifthenelse{\boolean{articletitles}}{\emph{{PYTHIA 6.4 physics and manual}},
  }{}\href{http://dx.doi.org/10.1088/1126-6708/2006/05/026}{JHEP \textbf{05}
  (2006) 026}, \href{http://arxiv.org/abs/hep-ph/0603175}{{\normalfont\ttfamily
  arXiv:hep-ph/0603175}}\relax
\mciteBstWouldAddEndPuncttrue
\mciteSetBstMidEndSepPunct{\mcitedefaultmidpunct}
{\mcitedefaultendpunct}{\mcitedefaultseppunct}\relax
\EndOfBibitem
\bibitem{Sjostrand:2007gs}
T.~Sj\"{o}strand, S.~Mrenna, and P.~Skands,
  \ifthenelse{\boolean{articletitles}}{\emph{{A brief introduction to PYTHIA
  8.1}}, }{}\href{http://dx.doi.org/10.1016/j.cpc.2008.01.036}{Comput.\ Phys.\
  Commun.\  \textbf{178} (2008) 852},
  \href{http://arxiv.org/abs/0710.3820}{{\normalfont\ttfamily
  arXiv:0710.3820}}\relax
\mciteBstWouldAddEndPuncttrue
\mciteSetBstMidEndSepPunct{\mcitedefaultmidpunct}
{\mcitedefaultendpunct}{\mcitedefaultseppunct}\relax
\EndOfBibitem
\bibitem{LHCb-PROC-2010-056}
I.~Belyaev {\em et~al.}, \ifthenelse{\boolean{articletitles}}{\emph{{Handling
  of the generation of primary events in Gauss, the LHCb simulation
  framework}}, }{}\href{http://dx.doi.org/10.1088/1742-6596/331/3/032047}{{J.\
  Phys.\ Conf.\ Ser.\ } \textbf{331} (2011) 032047}\relax
\mciteBstWouldAddEndPuncttrue
\mciteSetBstMidEndSepPunct{\mcitedefaultmidpunct}
{\mcitedefaultendpunct}{\mcitedefaultseppunct}\relax
\EndOfBibitem
\bibitem{Lange:2001uf}
D.~J. Lange, \ifthenelse{\boolean{articletitles}}{\emph{{The EvtGen particle
  decay simulation package}},
  }{}\href{http://dx.doi.org/10.1016/S0168-9002(01)00089-4}{Nucl.\ Instrum.\
  Meth.\  \textbf{A462} (2001) 152}\relax
\mciteBstWouldAddEndPuncttrue
\mciteSetBstMidEndSepPunct{\mcitedefaultmidpunct}
{\mcitedefaultendpunct}{\mcitedefaultseppunct}\relax
\EndOfBibitem
\bibitem{Golonka:2005pn}
P.~Golonka and Z.~Was, \ifthenelse{\boolean{articletitles}}{\emph{{PHOTOS Monte
  Carlo: A precision tool for QED corrections in $Z$ and $W$ decays}},
  }{}\href{http://dx.doi.org/10.1140/epjc/s2005-02396-4}{Eur.\ Phys.\ J.\
  \textbf{C45} (2006) 97},
  \href{http://arxiv.org/abs/hep-ph/0506026}{{\normalfont\ttfamily
  arXiv:hep-ph/0506026}}\relax
\mciteBstWouldAddEndPuncttrue
\mciteSetBstMidEndSepPunct{\mcitedefaultmidpunct}
{\mcitedefaultendpunct}{\mcitedefaultseppunct}\relax
\EndOfBibitem
\bibitem{Allison:2006ve}
Geant4 collaboration, J.~Allison {\em et~al.},
  \ifthenelse{\boolean{articletitles}}{\emph{{Geant4 developments and
  applications}}, }{}\href{http://dx.doi.org/10.1109/TNS.2006.869826}{IEEE
  Trans.\ Nucl.\ Sci.\  \textbf{53} (2006) 270}\relax
\mciteBstWouldAddEndPuncttrue
\mciteSetBstMidEndSepPunct{\mcitedefaultmidpunct}
{\mcitedefaultendpunct}{\mcitedefaultseppunct}\relax
\EndOfBibitem
\bibitem{Agostinelli:2002hh}
Geant4 collaboration, S.~Agostinelli {\em et~al.},
  \ifthenelse{\boolean{articletitles}}{\emph{{Geant4: A simulation toolkit}},
  }{}\href{http://dx.doi.org/10.1016/S0168-9002(03)01368-8}{Nucl.\ Instrum.\
  Meth.\  \textbf{A506} (2003) 250}\relax
\mciteBstWouldAddEndPuncttrue
\mciteSetBstMidEndSepPunct{\mcitedefaultmidpunct}
{\mcitedefaultendpunct}{\mcitedefaultseppunct}\relax
\EndOfBibitem
\bibitem{LHCb-PROC-2011-006}
M.~Clemencic {\em et~al.}, \ifthenelse{\boolean{articletitles}}{\emph{{The
  \lhcb simulation application, Gauss: Design, evolution and experience}},
  }{}\href{http://dx.doi.org/10.1088/1742-6596/331/3/032023}{{J.\ Phys.\ Conf.\
  Ser.\ } \textbf{331} (2011) 032023}\relax
\mciteBstWouldAddEndPuncttrue
\mciteSetBstMidEndSepPunct{\mcitedefaultmidpunct}
{\mcitedefaultendpunct}{\mcitedefaultseppunct}\relax
\EndOfBibitem
\bibitem{LHCb-DP-2012-004}
R.~Aaij {\em et~al.}, \ifthenelse{\boolean{articletitles}}{\emph{{The \lhcb
  trigger and its performance in 2011}},
  }{}\href{http://dx.doi.org/10.1088/1748-0221/8/04/P04022}{JINST \textbf{8}
  (2013) P04022}, \href{http://arxiv.org/abs/1211.3055}{{\normalfont\ttfamily
  arXiv:1211.3055}}\relax
\mciteBstWouldAddEndPuncttrue
\mciteSetBstMidEndSepPunct{\mcitedefaultmidpunct}
{\mcitedefaultendpunct}{\mcitedefaultseppunct}\relax
\EndOfBibitem
\bibitem{BBDT}
V.~V. Gligorov and M.~Williams,
  \ifthenelse{\boolean{articletitles}}{\emph{{Efficient, reliable and fast
  high-level triggering using a bonsai boosted decision tree}},
  }{}\href{http://dx.doi.org/10.1088/1748-0221/8/02/P02013}{JINST \textbf{8}
  (2013) P02013}, \href{http://arxiv.org/abs/1210.6861}{{\normalfont\ttfamily
  arXiv:1210.6861}}\relax
\mciteBstWouldAddEndPuncttrue
\mciteSetBstMidEndSepPunct{\mcitedefaultmidpunct}
{\mcitedefaultendpunct}{\mcitedefaultseppunct}\relax
\EndOfBibitem
\bibitem{Feindt2006190}
M.~Feindt and U.~Kerzel, \ifthenelse{\boolean{articletitles}}{\emph{{The
  NeuroBayes neural network package}},
  }{}\href{http://dx.doi.org/10.1016/j.nima.2005.11.166}{Nucl.\ Instrum.\
  Meth.\  \textbf{A559} (2006) 190}\relax
\mciteBstWouldAddEndPuncttrue
\mciteSetBstMidEndSepPunct{\mcitedefaultmidpunct}
{\mcitedefaultendpunct}{\mcitedefaultseppunct}\relax
\EndOfBibitem
\bibitem{LHCb-PAPER-2012-001}
LHCb collaboration, R.~Aaij {\em et~al.},
  \ifthenelse{\boolean{articletitles}}{\emph{{Observation of $\CP$ violation in
  $\Bpm\to\D\Kpm$ decays}},
  }{}\href{http://dx.doi.org/10.1016/j.physletb.2012.04.060}{Phys.\ Lett.\
  \textbf{B712} (2012) 203}, Erratum
  \href{http://dx.doi.org/10.1016/j.physletb.2012.05.060}{ibid.\
  \textbf{B713} (2012) 351},
  \href{http://arxiv.org/abs/1203.3662}{{\normalfont\ttfamily
  arXiv:1203.3662}}\relax
\mciteBstWouldAddEndPuncttrue
\mciteSetBstMidEndSepPunct{\mcitedefaultmidpunct}
{\mcitedefaultendpunct}{\mcitedefaultseppunct}\relax
\EndOfBibitem
\bibitem{Pivk:2004ty}
M.~Pivk and F.~R. Le~Diberder,
  \ifthenelse{\boolean{articletitles}}{\emph{{sPlot: A statistical tool to
  unfold data distributions}},
  }{}\href{http://dx.doi.org/10.1016/j.nima.2005.08.106}{Nucl.\ Instrum.\
  Meth.\  \textbf{A555} (2005) 356},
  \href{http://arxiv.org/abs/physics/0402083}{{\normalfont\ttfamily
  arXiv:physics/0402083}}\relax
\mciteBstWouldAddEndPuncttrue
\mciteSetBstMidEndSepPunct{\mcitedefaultmidpunct}
{\mcitedefaultendpunct}{\mcitedefaultseppunct}\relax
\EndOfBibitem
\bibitem{Punzi:2003bu}
G.~Punzi, \ifthenelse{\boolean{articletitles}}{\emph{{Sensitivity of searches
  for new signals and its optimization}}, }{} in {\em Statistical Problems in
  Particle Physics, Astrophysics, and Cosmology} (L.~{Lyons}, R.~{Mount}, and
  R.~{Reitmeyer}, eds.), p.~79, 2003.
\newblock \href{http://arxiv.org/abs/physics/0308063}{{\normalfont\ttfamily
  arXiv:physics/0308063}}\relax
\mciteBstWouldAddEndPuncttrue
\mciteSetBstMidEndSepPunct{\mcitedefaultmidpunct}
{\mcitedefaultendpunct}{\mcitedefaultseppunct}\relax
\EndOfBibitem
\bibitem{LHCb-DP-2012-003}
M.~Adinolfi {\em et~al.},
  \ifthenelse{\boolean{articletitles}}{\emph{{Performance of the \lhcb RICH
  detector at the LHC}},
  }{}\href{http://dx.doi.org/10.1140/epjc/s10052-013-2431-9}{Eur.\ Phys.\ J.\
  \textbf{C73} (2013) 2431},
  \href{http://arxiv.org/abs/1211.6759}{{\normalfont\ttfamily
  arXiv:1211.6759}}\relax
\mciteBstWouldAddEndPuncttrue
\mciteSetBstMidEndSepPunct{\mcitedefaultmidpunct}
{\mcitedefaultendpunct}{\mcitedefaultseppunct}\relax
\EndOfBibitem
\bibitem{Skwarnicki:1986xj}
T.~Skwarnicki, {\em {A study of the radiative cascade transitions between the
  Upsilon-prime and Upsilon resonances}}, PhD thesis, Institute of Nuclear
  Physics, Krakow, 1986,
  {\href{http://inspirehep.net/record/230779/}{DESY-F31-86-02}}\relax
\mciteBstWouldAddEndPuncttrue
\mciteSetBstMidEndSepPunct{\mcitedefaultmidpunct}
{\mcitedefaultendpunct}{\mcitedefaultseppunct}\relax
\EndOfBibitem
\bibitem{LHCb-PAPER-2014-048}
LHCb collaboration, R.~Aaij {\em et~al.},
  \ifthenelse{\boolean{articletitles}}{\emph{{Precision measurement of the mass
  and lifetime of the $\Xibm$ baryon}},
  }{}\href{http://dx.doi.org/10.1103/PhysRevLett.113.242002}{Phys.\ Rev.\
  Lett.\  \textbf{113} (2014) 242002},
  \href{http://arxiv.org/abs/1409.8568}{{\normalfont\ttfamily
  arXiv:1409.8568}}\relax
\mciteBstWouldAddEndPuncttrue
\mciteSetBstMidEndSepPunct{\mcitedefaultmidpunct}
{\mcitedefaultendpunct}{\mcitedefaultseppunct}\relax
\EndOfBibitem
\bibitem{LHCb-PAPER-2016-008}
LHCb collaboration, R.~Aaij {\em et~al.},
  \ifthenelse{\boolean{articletitles}}{\emph{{Measurements of the mass and
  lifetime of the $\Omegab$ baryon}},
  }{}\href{http://dx.doi.org/10.1103/PhysRevD.93.092007}{Phys.\ Rev.\
  \textbf{D93} (2016) 092007},
  \href{http://arxiv.org/abs/1604.01412}{{\normalfont\ttfamily
  arXiv:1604.01412}}\relax
\mciteBstWouldAddEndPuncttrue
\mciteSetBstMidEndSepPunct{\mcitedefaultmidpunct}
{\mcitedefaultendpunct}{\mcitedefaultseppunct}\relax
\EndOfBibitem
\bibitem{HFAG}
Heavy Flavor Averaging Group, Y.~Amhis {\em et~al.},
  \ifthenelse{\boolean{articletitles}}{\emph{{Averages of $b$-hadron,
  $c$-hadron, and $\tau$-lepton properties as of summer 2014}},
  }{}\href{http://arxiv.org/abs/1412.7515}{{\normalfont\ttfamily
  arXiv:1412.7515}}, {updated results and plots available at
  \href{http://www.slac.stanford.edu/xorg/hfag/}{{\texttt{http://www.slac.stanford.edu/xorg/hfag/}}}}\relax
\mciteBstWouldAddEndPuncttrue
\mciteSetBstMidEndSepPunct{\mcitedefaultmidpunct}
{\mcitedefaultendpunct}{\mcitedefaultseppunct}\relax
\EndOfBibitem
\bibitem{Albrecht:1990am}
ARGUS collaboration, H.~Albrecht {\em et~al.},
  \ifthenelse{\boolean{articletitles}}{\emph{{Search for hadronic $b \to u$
  decays}}, }{}\href{http://dx.doi.org/10.1016/0370-2693(90)91293-K}{Phys.\
  Lett.\  \textbf{B241} (1990) 278}\relax
\mciteBstWouldAddEndPuncttrue
\mciteSetBstMidEndSepPunct{\mcitedefaultmidpunct}
{\mcitedefaultendpunct}{\mcitedefaultseppunct}\relax
\EndOfBibitem
\bibitem{PDG2014}
Particle Data Group, K.~A. Olive {\em et~al.},
  \ifthenelse{\boolean{articletitles}}{\emph{{\href{http://pdg.lbl.gov/}{Review
  of particle physics}}},
  }{}\href{http://dx.doi.org/10.1088/1674-1137/38/9/090001}{Chin.\ Phys.\
  \textbf{C38} (2014) 090001}, {and 2015 update}\relax
\mciteBstWouldAddEndPuncttrue
\mciteSetBstMidEndSepPunct{\mcitedefaultmidpunct}
{\mcitedefaultendpunct}{\mcitedefaultseppunct}\relax
\EndOfBibitem
\bibitem{Anderlini:2202412}
L.~Anderlini {\em et~al.}, \ifthenelse{\boolean{articletitles}}{\emph{{The
  PIDCalib package}}, }{}
  \href{http://cdsweb.cern.ch/search?p=LHCb-PUB-2016-021&f=reportnumber&action_search=Search&c=LHCb+Notes}
  {LHCb-PUB-2016-021}\relax
\mciteBstWouldAddEndPuncttrue
\mciteSetBstMidEndSepPunct{\mcitedefaultmidpunct}
{\mcitedefaultendpunct}{\mcitedefaultseppunct}\relax
\EndOfBibitem
\bibitem{LHCb-PAPER-2012-018}
LHCb collaboration, R.~Aaij {\em et~al.},
  \ifthenelse{\boolean{articletitles}}{\emph{{Observation of $\Bz\to\Dzb\Kp\Km$
  and evidence for $\Bs\to\Dzb\Kp\Km$}},
  }{}\href{http://dx.doi.org/10.1103/PhysRevLett.109.131801}{Phys.\ Rev.\
  Lett.\  \textbf{109} (2012) 131801},
  \href{http://arxiv.org/abs/1207.5991}{{\normalfont\ttfamily
  arXiv:1207.5991}}\relax
\mciteBstWouldAddEndPuncttrue
\mciteSetBstMidEndSepPunct{\mcitedefaultmidpunct}
{\mcitedefaultendpunct}{\mcitedefaultseppunct}\relax
\EndOfBibitem
\bibitem{LHCb-PAPER-2015-029}
LHCb collaboration, R.~Aaij {\em et~al.},
  \ifthenelse{\boolean{articletitles}}{\emph{{Observation of $\jpsi\proton$
  resonances consistent with pentaquark states in $\Lb\to\jpsi\proton\Km$
  decays}}, }{}\href{http://dx.doi.org/10.1103/PhysRevLett.115.072001}{Phys.\
  Rev.\ Lett.\  \textbf{115} (2015) 072001},
  \href{http://arxiv.org/abs/1507.03414}{{\normalfont\ttfamily
  arXiv:1507.03414}}\relax
\mciteBstWouldAddEndPuncttrue
\mciteSetBstMidEndSepPunct{\mcitedefaultmidpunct}
{\mcitedefaultendpunct}{\mcitedefaultseppunct}\relax
\EndOfBibitem
\bibitem{LHCb-PAPER-2014-036}
LHCb collaboration, R.~Aaij {\em et~al.},
  \ifthenelse{\boolean{articletitles}}{\emph{{Dalitz plot analysis of
  $\Bs\to\Dzb\Km\pip$ decays}},
  }{}\href{http://dx.doi.org/10.1103/PhysRevD.90.072003}{Phys.\ Rev.\
  \textbf{D90} (2014) 072003},
  \href{http://arxiv.org/abs/1407.7712}{{\normalfont\ttfamily
  arXiv:1407.7712}}\relax
\mciteBstWouldAddEndPuncttrue
\mciteSetBstMidEndSepPunct{\mcitedefaultmidpunct}
{\mcitedefaultendpunct}{\mcitedefaultseppunct}\relax
\EndOfBibitem
\end{mcitethebibliography}
\ifx\mcitethebibliography\mciteundefinedmacro
\PackageError{LHCb.bst}{mciteplus.sty has not been loaded}
{This bibstyle requires the use of the mciteplus package.}\fi
\providecommand{\href}[2]{#2}

\newpage
\centerline{\large\bf LHCb collaboration}
\begin{flushleft}
\small
R.~Aaij$^{40}$,
B.~Adeva$^{39}$,
M.~Adinolfi$^{48}$,
Z.~Ajaltouni$^{5}$,
S.~Akar$^{59}$,
J.~Albrecht$^{10}$,
F.~Alessio$^{40}$,
M.~Alexander$^{53}$,
S.~Ali$^{43}$,
G.~Alkhazov$^{31}$
% ...

{\footnotesize \it
% ...
$ ^{5}$Clermont Universit{\'e}, Universit{\'e} Blaise Pascal, CNRS/IN2P3, LPC, Clermont-Ferrand, France\\
% ...
$ ^{10}$Fakult{\"a}t Physik, Technische Universit{\"a}t Dortmund, Dortmund, Germany\\
% ...
$ ^{31}$Petersburg Nuclear Physics Institute (PNPI), Gatchina, Russia\\
% ...
$ ^{39}$Universidad de Santiago de Compostela, Santiago de Compostela, Spain\\
$ ^{40}$European Organization for Nuclear Research (CERN), Geneva, Switzerland\\
% ...
$ ^{43}$Nikhef National Institute for Subatomic Physics, Amsterdam, The Netherlands\\
% ...
$ ^{48}$H.H. Wills Physics Laboratory, University of Bristol, Bristol, United Kingdom\\
% ...
$ ^{53}$School of Physics and Astronomy, University of Glasgow, Glasgow, United Kingdom\\
% ...
$ ^{59}$University of Cincinnati, Cincinnati, OH, United States\\
% ...
}
\end{flushleft}

\end{document}